\documentclass[twocolumn,showpacs,preprintnumbers,amsmath,amssymb]{revtex4}
\usepackage{epsfig}
\usepackage{amsmath}

\begin{document}

\title{The Web of Connections between Tourism Companies in Elba: Structure and Dynamics}

\author{Luciano da Fontoura Costa}
\affiliation{Institute of Physics at S\~ao Carlos, University of
S\~ao Paulo, P.O. Box 369, S\~ao Carlos, S\~ao Paulo, 13560-970
Brazil. Email: luciano@ifsc.usp.br}

\author{Rodolfo Baggio}
\affiliation{Master in Economics and Tourism, 
Bocconi University, via Sarfatti, 25, 
20136 Milan, Italy. Email: rodolfo.baggio@unibocconi.it}

\date{24th Mar 2008}

\begin{abstract}
Tourism destination networks are amongst the most complex dynamical
systems, involving a myriad of human-made and natural resources. In
this work we report a complex network-based systematic analysis of the
Elba (Italy) tourism destination network, including the
characterization of its structure in terms of a set of several
traditional measurements, the investigation of its modularity, as well
as its comprehensive study in terms of the recently reported
superedges approach.  In particular, structural (the number of paths
of distinct lengths between pairs of nodes, as well as the number of
reachable companies) and dynamical features (transition probabilities
and the inward/outward activations and accessibilities) are measured
and analyzed, leading to a series of important findings related to the
interactions between tourism companies.  Among the several reported
results, it is shown that the type and size of the companies influence
strongly their respective activations and accessibilities, while their
geographical position does not seem to matter.  It is also shown that
the Elba tourism network is largely fragmented and heterogeneous, so
that it could benefit from increased integration.

\vspace{0.5cm}


\end{abstract}

\maketitle

\vspace{0.5cm}

\emph{`Make voyages! Attempt them! - there's nothing else...'
(T. Williams)}

\section{Introduction} 

Tourism is today probably the largest economic sector of the world
economy. This economic system has fairly indefinite boundaries and
comprises a wide diversity of organizations offering various products
and services which exhibit very little
homogeneity~\cite{Vanhove:2005}.  A tourism destination, loosely
defined as the goal of a traveler, is considered a fundamental unit
of analysis for the understanding of the whole tourism
sector~\cite{Jafari:2000}. From a socio-economic viewpoint it consists
of a number of companies and organizations (public and private) who
manage different attractions and services to be offered a
visitor~\cite{Baggio:2008}. A tourism destination is a complex
adaptive system sharing many (if not all) of the characteristics
usually associated with it: non-linear relationships among the
components (companies and organizations), self organization and
emergence of organizational structures, robustness to external
shocks~\cite{Baggio:2008, Faulkner:1997}. The dynamic set of
relationships which form the connective tissue holding together the
system's elements suggests a network approach to be indispensable for
the understanding of a tourism destination. Several authors have used
this perspective, mostly at a qualitative level~\cite{Lazzeretti:2006,
Morrison:2004, Pavlovich:2003}. Only a few, however, have started
applying quantitative methods and tools of this area in order to
improve our knowledge of the structure and the dynamic behavior of a
tourism system~\cite{Scott:2008, Scott_b:2008, Tallinucci:2006,
Bramwell:2000, Baggio:2007}.  Today, more than ever, strong
international competition induces an imperative to innovate to remain
competitive. Many authors recognize that a pre-requisite for
innovation is the capability to cooperate and collaborate efficiently
and effectively. Tourism, more than most economic sectors, involves
the development of formal and informal collaborations, partnerships
and networks, and, rather obviously, the understanding of the patterns
of linkages among the destination components and the assessment of the
system's structure are crucial points~\cite{Scott_b:2008,
Bramwell:2000, Cooper:2006, Scott:2007}.  Not less important is the
effective access to local information and knowledge maintained by each
participant of this intricate system.

With its origins going back to Flory~\cite{Flory} and
Erd\H{o}s-R\'enyi~\cite{Erdos_Reny:1959} works on random graphs, the
area of complex networks~\cite{Albert_Barab:2002, Dorogov_Mendes:2002,
Newman:2003, Boccaletti:2006, Costa_surv:2007} has established itself
as one of the most dynamic and exciting alternatives for representing
the structure and dynamics of the most diverse natural and human-made
complex systems.  One of the main reasons for the growing popularity
and success of complex networks investigations consists in its
generality for representing and modeling virtually any system composed
of discrete parts (e.g.~\cite{Costa_appl:2008}), encompassing from
protein-protein interaction (e.g.~\cite{Jeong:2001}) to scientific
collaboration (e.g.~\cite{Lehmann:2003}).  In addition, by
representing any type of connectivity, complex networks are
intrinsically suited for the investigation of more general types of
dynamics, as opposite to the consideration of regular lattices
adopted by the majority of previous works.  As a matter of fact,
growing attention in complex networks research has been focused on
investigations of relationships between the structure and dynamics
(e.g.~\cite{Newman:2003, Boccaletti:2006, Dorogov_Mendes:2002}).
Three of the most important subjects currently pursued by complex
network scientists correspond to: (i) the characterization of the
structure of complex systems by using several topological measurements
(e.g.~\cite{Costa_surv:2007}); (ii) the investigation of the
modularity (i.e. community finding) of complex networks; and (iii)
studies of the relationship between structure and dynamics of complex
systems.

The current work reports a complex network approach to the
comprehensive investigation of the complex system corresponding to the
tourism destinations in the Island of Elba.  Each tourism agent is
represented as a node, while the relationships between such agents are
expressed by undirected edges.  Our investigation is considered from
the perspective of all the three main approaches identified above,
namely structural characterization in terms of several measurements,
identification of communities, and investigation of the relationship
between structure and function by using the \emph{superedges} concept
introduced recently~\cite{Costa_superedges:2008}.  The work starts by
describing the construction of the tourism destination network and
proceeds by reporting the characterization of its structure and
modularity, followed by the superedges approach to the
structure:dynamics investigation.

\section{The Tourism Destination Network}

The destination analyzed here is the island of Elba, Italy. It belongs
to the Tuscany Archipelago National Park (located in the central
Thyrrenian sea) and is the third Italian island. It is an important
environmental resource and a significant contributor to the country's
economy. Almost 500,000 tourists spend some 3 million nights per year
in several hundred accommodation establishments. Elba is considered a
mature tourism destination with a long history and which has gone
through a number of different expansion and reorganization cycles. The
great majority of the stakeholders are small and medium sized
companies, mostly family-run. Several associations and consortia
operate on the island and try to recommend and develop different types
of collaboration programs in an attempt to overcome the excessive
'independence' of the local companies~\cite{Pechlaner:2003,
Tallinucci:2006}.  The destination network was built in the following
way. The core tourism companies and associations operating at Elba are
considered the nodes of a network whose ties are the relationships
among them. According to the local tourism board, the list of
companies comprises 1028 items. the links reflect 'business' relations
between organizations. They were collected by consulting publicly
available sources such as associations listings, management board
compositions, catalogs of travel agencies, marketing leaflets and
brochures, official corporate records (to assess the belonging to
industrial groups). These data were then verified with a series of
in-depth interviews to 'knowledgeable informants': director of tourism
board, directors of associations, tourism consultants etc. This
triangulation~\cite{Olsen:2004} allowed to validate existing linkages
and uncover others.  The so-obtained network can be reasonably
estimated to be nearly $90\%$ complete.

Finally, based on the information available, all the nodes were
recorded along with their belonging to a specific type of business (8
types: e.g. hotels, travel agencies, associations etc.), geographical
location (9 areas reflecting Elba's municipalities) and size (small,
medium, large, estimated on the real size of the company). Overall, 8
different types, 9 geographical areas and 3 sizes are
present. Table~\ref{tab:type} shows the different node groupings
according to the three main classifications.

\begin{table}[h]
	\caption{Types of Elba network operators.} \vspace{0.5cm}
	\begin{tabular}{|c|l|c|l|c|l|} \hline
	\multicolumn{2}{|c|}{Type of business} & \multicolumn{2}{|c|}{Geography} & \multicolumn{2}{|c|}{Size} \\ \hline
ID&Type &ID&Location &ID&Size \\ \hline
1&Associations &1& Porto Azzurro &1& Large \\ \hline
2&Cultural resources&2& Portoferraio &2& Medium\\ \hline
3&Food and Beverage & 3& Capoliveri &3&Small \\ \hline
4&Hospitality & 4&Rio Marina && \\ \hline
5&Intermediaries &5& Rio nell'Elba & &\\ \hline
6&Public Organizations &6& Campo nell'Elba & &\\ \hline
7&Transports/Rentals &7&   Marciana & &\\ \hline
8&Other services &8& Marciana Marina && \\ \hline
& &9&All island && \\ \hline
	\end{tabular}
	\label{tab:type}
\end{table}

\section{Characterization of the Tourism Destination Network}

Complex network analysis methods were used to assess the topological
characteristics of the system (for definitions and formulas see for
example~\cite{Costa_surv:2007}). The obtained measurements, shown in
Table~\ref{tab:meas}, were calculated by using available software
packages (Pajek, Ucinet) complemented by some Matlab programs
developed by one of the authors. Degree distribution scaling exponents
are calculated according to Clauset et al.~\cite{Clauset:2007}. 

\begin{table}[b]
\caption{Values of several measurements 
calculated for the Elba tourism destination network.}  \vspace{0.5cm}
\begin{tabular}{|l|c|}  \hline
Metric	& Value \\  \hline
Number of nodes	& 1028 \\  \hline 
Number of edges	& 1642 \\  \hline
Density	& 0.003 \\  \hline
Disconnected nodes	& $37\%$ \\  \hline
Diameter	&  8 \\  \hline
Average path length	&  3.16 \\  \hline
Clustering coefficient	&  0.050 \\  \hline
Proximity ratio	&  34.10 \\  \hline
Average degree	&  3.19 \\  \hline
Average closeness	& 0.121 \\  \hline
Average betweenness	& 0.001 \\  \hline
Global efficiency	& 0.131 \\  \hline
Local efficiency	& 0.062 \\  \hline
Assortativity coefficient	& -0.164 \\  \hline
Degree distribution exponent	& 2.32 \\  \hline
\end{tabular}~\label{tab:meas}
\end{table}

A modularity analysis was also performed. The modularity
index~\cite{Newman_Girvan:2004} is shown in Table~\ref{tab:modul} with
respect to several networks. The network nodes were divided into
groups according to their typology as tourism operators and to the
geographical location inside the island. Moreover, the method proposed
by Clauset et al.~\cite{Clauset:2004} was used to identify
algorithmically the community structure of the network (CNM in the
first column). As a comparison, the last row gives the values
calculated (CNM) for a network of the same order as the Elba network
with a random distribution of links (values are averages over 10
realizations). It is important to note that the groups identified by using
this method (CNM) are different in number and composition from the
others (geography and type). In order to better evaluate the different
results, the last column of the table contains the average modularity
over the groups (modularity/number of groups).
It must be added here that the majority of the modules identified by the CNM 
algorithm fall within the resolution limits set by Fortunato and Barth\'elemy~\cite{Fortunato:2007}, thus suggesting the existence of a finer structure. However, for the objectives of the present investigation, the analysis was not conducted any further.

\begin{table*}
\caption{Modularity analysis of the Elba network.}  \vspace{0.5cm}
\begin{tabular}{|l|c|c|c|c|}  \hline
Grouping & Number of groups & Modularity & Average Modularity  \\ \hline
 Geography & 9 & 0.047 & 0.0052 \\ \hline
 Type & 8 & -0.255 &  -0.0319  \\ \hline
 CNM  & 11  & 0.396 & 0.0360 \\ \hline
 CNM (random) & 23 & 0.606 & 0.0263 \\ \hline
\end{tabular}~\label{tab:modul}
\end{table*}

All in all, the main findings of the analysis of the main structural
characteristics of the Elba network can be summarized as follows:

\begin{list}
\centering
\item $\bullet$ The network shows a scale-free topology (power-law 
behavior of the degree distribution) which is consistent with that
generally ascribed to many artificial and natural complex networks,
moreover it shows a certain degree of small-worldness as shown by the
proximity ratio;
\item $\bullet$ The general connectivity is very low (link density) 
with a very large proportion of disconnected elements;
\item $\bullet$ Clustering is quite limited, as is the efficiency, 
both at a local and global level;
\item $\bullet$ Assortativity is negative, contrary to the general 
findings that show positive values for social and economic networks;
\item $\bullet$ The modularity is generally very low. In one case, 
by type of business, it is negative. This means that companies
(e.g. an hotel) tend to connect with some other company which is not
of the same business.
\end{list}

These results provide quantitative evidence in favor of recognizing
that the community of Elban tourism operators is very fragmented in
nature. There appears to be little incentive to group or cluster in a
cooperative or collaborative manner as evidenced by considering the
clustering and assortativity characteristics. The study of modularity
further confirms this finding. Sadly, this is a quite common
phenomenon in many tourism destinations. Similar tourism operators
dislike each other more than trying to combine their resources (at
least some) to better cope with the market. A significant example of
strong competition. These conditions are also problematic for an
efficient flow of information and knowledge through the social system,
and this may affect its capabilities for innovation and future
competitiveness. These considerations are in general agreement with
previous studies performed by using more traditional qualitative
techniques~\cite{Pechlaner:2003, Tallinucci:2006}.

\section{The Superedges Approach to Structure and Dynamics}

\begin{figure}[htb]
  \vspace{0.3cm} \begin{center}
  \includegraphics[width=0.85\linewidth]{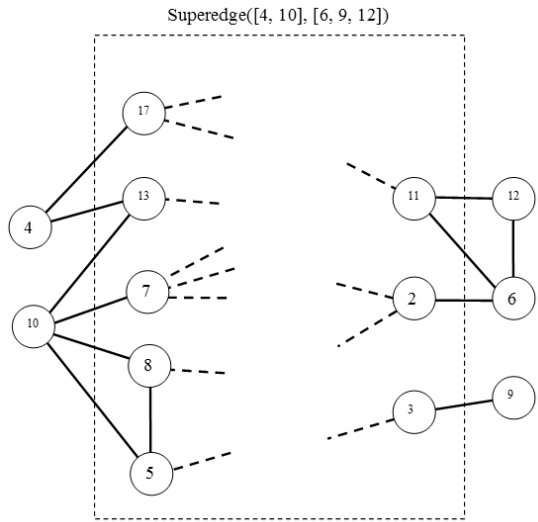} \\
  \caption{Illustration of the superedge approach.  Two sets of nodes
             are first selected as input (in this example $[4, 10]$)
             and output ($6, 9, 12$).  Then the connectivity between
             these two sets is quantified in terms of the number of
             paths of different lengths between those sets.  The
             dynamics induced over the output nodes implied by stimulation
             of the input set (e.g. liberation of moving agents at those
             nodes) is also quantified in terms of a set of respective
             measurements.  Finally, the structural and dynamical 
             features are investigated for the identification of possible
             relationships (e.g. correlations).
  }~\label{fig:superedges} \end{center}
\end{figure}

The above characterization of the tourism network was restricted to
topological features.  Because real tourism is a dynamical process, it
becomes important to consider also the study of its possible
\emph{dynamics}.  A comprehensive approach to investigating and relating
structure and dynamics in complex systems, herein called the 
\emph{superedges approach}, was reported 
recently~\cite{Costa_superedges:2008}.  This approach is founded on
the treatment of a complex network as a dynamical system, considering
subsets of nodes as respective input and output, as illustrated in
Figure~\ref{fig:superedges}.

For each specific choice of input and output, the connectivity between
these two sets of nodes is characterized comprehensively in terms of
the number of paths with distinct lengths extending from the input to
the output (other measurements, such as the properties of the nodes
along the identified paths, can be also considered).  Such an approach
provides a substantially more comprehensive characterization of the
topology of the connections than the typically used measurements of
node degree and shortest paths (actually, the shortest paths are
naturally incorporated into the superedges approach).  Observe also
that the obtained distribution of paths yields a comprehensive
characterization of the \emph{structure} of the network with respect
to the chosen input and output sets.  Now, the dynamics of this
configuration can also be studied by adopting a specific type of
dynamics (e.g. traditional random walks, self-avoiding random walks,
Ising, or integrate-and-fire neuronal models).  The choice of the
specific type of dynamics should reflect the nature of the problem
under investigation and the respective questions which are being
posed.  Having chosen the specific dynamical rules, the network is
stimulated from the input set of nodes (e.g. by liberating moving
agents), while the respective effect over the output set is monitored
and characterized in terms of a set of measurements.  At the end of
such simulations, we have a set of structural measurements and a set
of measurements of the dynamics, which can then be related for
instance by using Pearson correlation
coefficient~\cite{Costa_surv:2007, Costa_superedges:2008}.

Therefore, the superedges approach to relating structure and dynamics
involves the following four basic steps: (i) selection of the input
and output sets of nodes; (ii) characterization of the connectivity
between these sets by considering the number of paths of distinct
lengths interconnecting those sets (other complementary measurements
can be used); (iii) stimulating some type of dynamics through the
input set and measuring the dynamics implied onto the output set; and
(iv) relating the obtained structural and dynamical features.  Though
the superedges approach is underlain by these main steps, the specific
decision about what measurement and dynamics to adopt are
intrinsically related to each specific problem.

\begin{figure}[htb]
  \vspace{0.3cm} \begin{center}
  \includegraphics[width=0.75\linewidth]{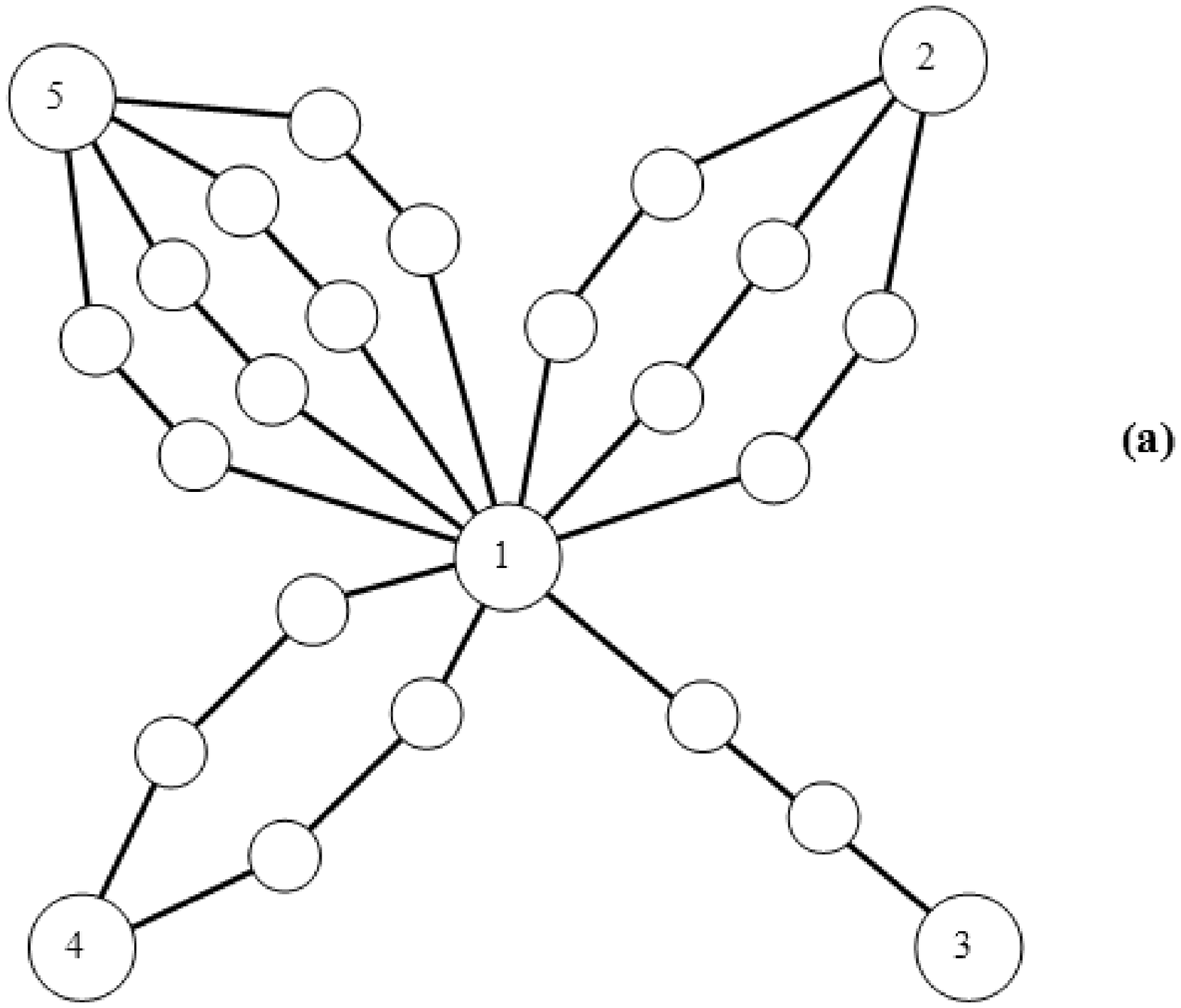} \\
  \includegraphics[width=0.75\linewidth]{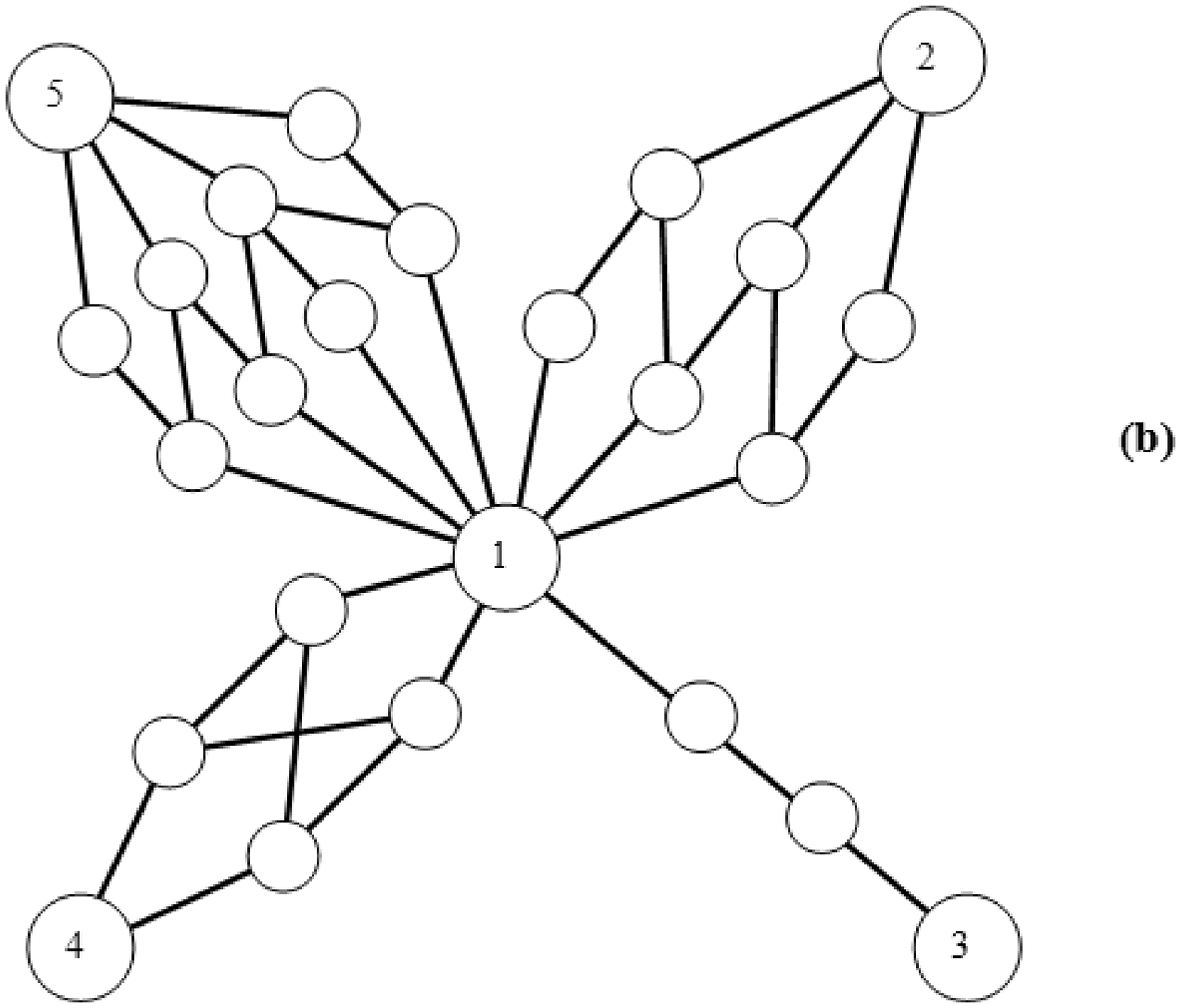} \\
  \includegraphics[width=0.75\linewidth]{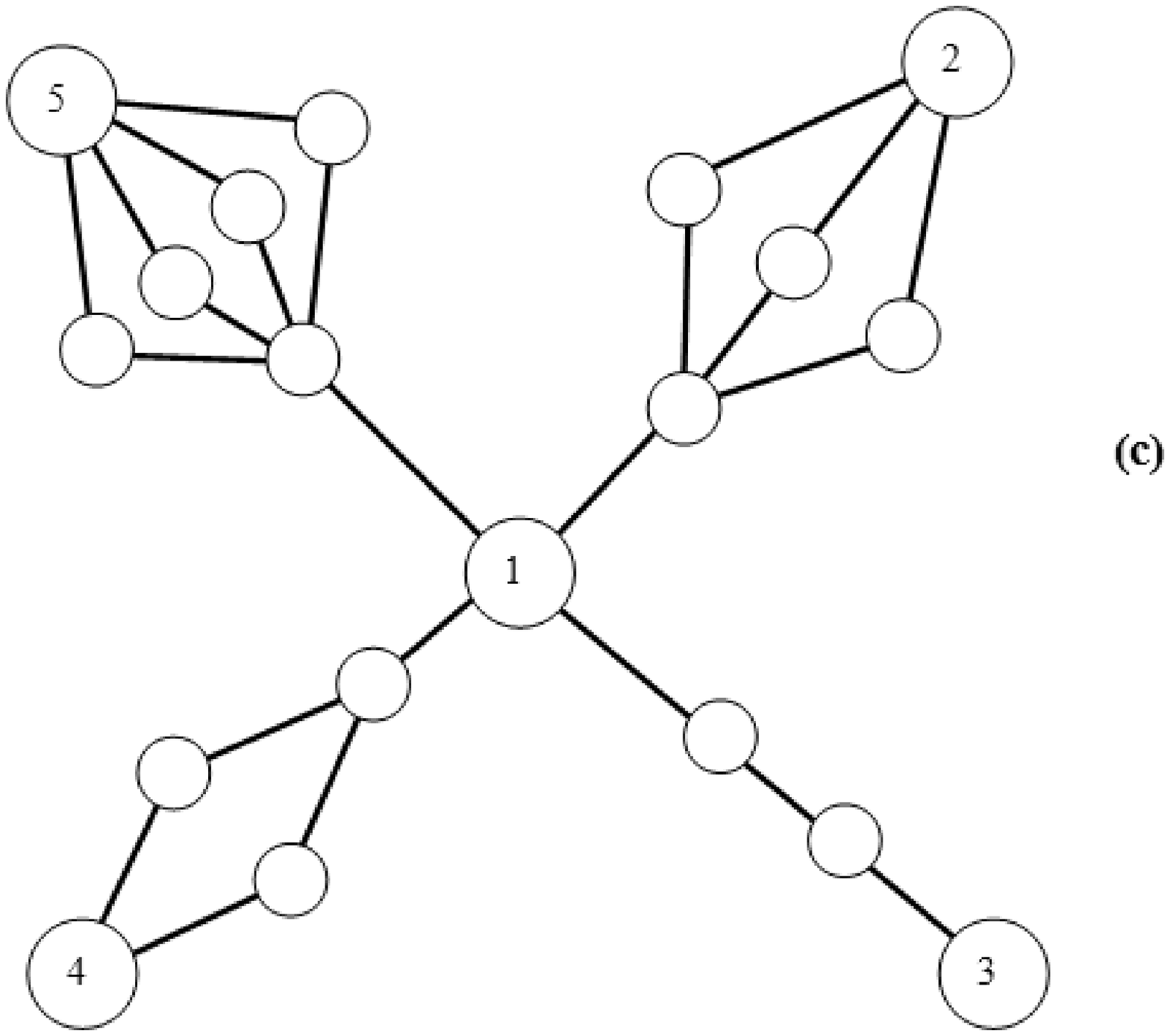} \\
  \caption{Three reference situations regarding the relationship between
           the number of paths and transition probabilities between
           pairs of nodes.  See text for explanation and discussion.
  }~\label{fig:probs} \end{center}
\end{figure}

The relationship between the number of paths of a given length $h$
between a node $i$ and a node $j$ (a topological measurement) and the
transition probability from node $i$ to node $j$ (a dynamical
measurement) deserves a preliminary discussion.
Figure~\ref{fig:probs}(a) shows a network where node $i=1$ is
connected to four other nodes (2 to 5) through several paths of length
3, more specifically: $Q_3(1,2) = 3$; $Q_3(1,3) = 1$; $Q_3(1,4) = 2$;
and $Q_3(1,5) = 4$.  The respective transition probabilities are:
$P_3(2,1) = 3/10$; $P_3(3,1) = 1/10$; $P_3(4,1) = 2/10$; and $P_3(5,1)
= 4/10$.  Therefore, in case of multiple independent paths as in
Figure~\ref{fig:probs}(a), the transition probability will be directly
proportional to the number of such paths.  Consider now the network
depicted in Figure~\ref{fig:probs}(b).  Now we have $Q_3(1,2) = 5$;
$Q_3(1,3) = 1$; $Q_3(1,4) = 4$; and $Q_3(1,5) = 7$, and $P_3(2,1) =
7/40$; $P_3(3,1) = 1/10$; $P_3(4,1) = 2/10$; and $P_3(5,1) \approx
0.22$.  So, it is clear that interdependencies between the paths tend
to break the correlation between the number of paths and the
transition probabilities.  Yet another possible situation is shown in
Figure~\ref{fig:probs}(c), for which $Q_3(1,2) = 3$; $Q_3(1,3) = 1$;
$Q_3(1,4) = 2$; and $Q_3(1,5) = 4$, and $P_3(2,1) = P_3(3,1) =
P_3(4,1) = P_3(5,1) = 1/4$, i.e. though we have the same number of
paths of length $2$ between nodes 1 to 5 as in
Figure~\ref{fig:probs}(a), the respective probabilities are completely
different, being determined by the first connections established by
node 1.  These three examples make it clear that there is no obvious
relationship or correlation between the number of paths and the
transition probabilities, which will ultimately be defined by
additional topological details of the networks as well as the specific
considered dynamics.  For such reasons, the comparison of these two
types of measurements presents good potential for characterizing and
understanding the complex systems under analysis.  For instance, a
strong correlation between the number of paths and transition suggests
the presence of independent paths such as in
Figure~\ref{fig:probs}(a).

Once the number of paths and transition probabilities associated to
each superedge have been calculated, it is possible to derive
additional measurements such as the activation and accessibility of
individual nodes~\cite{Costa_superedges:2008}.  More specifically, the
\emph{inward and outward activations} of a node $i$ at length $h$ are
defined respectively as:

\begin{eqnarray}
  Act\_in_h(i) =  \frac{\sum_{j \in \Omega} P_h(i,j)}{N-1} \nonumber \\
  Act\_out_h(i) =  \sum_{j \in \Omega} P_h(j,i)   \nonumber 
\end{eqnarray}

where $\Omega$ is the set of all nodes different from $i$.  The inward
and outward activations express the intensity of accesses into and
from each network node resulting from the respective topological and
dynamical properties.  In other words, the higher the inward
accessibility of a node, the larger the number of accesses that node
will receive.  A similar interpretation holds for the outward
activation, except that this value tends to become smaller than 1
because of the termination of paths at lengths smaller than the
current $h$ (branches leading to `dead-ends').  So, though the outward
activation generally tends to diminish with $h$ in a finite network,
the way in which such a decrease takes place depends on the topology
and dynamics in the network.

The \emph{inward and outward accessibilities} of a node $i$ are
respectively defined as

\begin{eqnarray}
  Acc\_in_h(i) =  \frac{exp(E_h(i, \Omega))}{N-1}  \nonumber \\
  Acc\_out_h(i) =  \frac{exp(E_h(\Omega, i))}{N-1}  \nonumber 
\end{eqnarray}

where $E_h(\Omega,i)$ and $E_h(\Omega,i)$ are the entropies of the
non-zero transition probabilities, defined as

\begin{eqnarray}
  E_h(i,\Omega) = -\sum_{j, P_h(i,j) \ne 0} \frac{P_h(i,j)}{N-1}
                       log \left( \frac{P_h(i,j)}{N-1} \right) \nonumber \\
  E_h(\Omega,i) = -\sum_{j, P_h(i,j) \ne 0}P_h(j,i) log(P_h(j,i)) \nonumber 
\end{eqnarray}

The inward and outward accessibilities quantify the effectiveness or
balance of accesses into and from node $i$ with respect to the other
nodes.  For instance, in case node $i$ is connected to $A$ nodes
through paths of length $h$, a high outward accessibility
$Acc\_out_h(i)$ implies that the $A$ nodes are accessed at similar
frequencies.  In addition, high outward accessibility also means that
all the $A$ nodes will be visited, in the average, in the shortest
period of time by self-avoiding random walks initiating at node $i$.
At the same time, a node $i$ with high inward accessibility for a
given $h$ will tend to be equally visited by all the nodes at that
distance.  While the activation quantifies the intensity of directed
accesses, the accessibility measures the equilibrium of such accesses.
Therefore, these measurements provide complementary characterization
of the dynamics of accesses unfolding in the network.

In the following section we report the configuration and application
of the superedges approach to the analysis of the relationship between
structure and function in tourism destination networks.

\section{The Superedges Analysis of Tourism Networks}

Basically, there are four main aspects to be specified while applying
the superedges approach: (a) the choice of the input and output sets;
(b) the choice of the measurements of the interconnection topology;
(c) the choice of the dynamics; and (d) the choice of the measurements
of the dynamics.  The selection of each of these aspects respectively
to tourism destination networks are explained in the following.

{\bf Input/Output Selection:} In order to obtain a systematic
investigation of the relationship between the involved companies, we
considering each possible node as input and output, implying a total
of $N(N-1)$ input-output configurations.  Therefore, we will be taking
into account all pairwise interactions between nodes.

{\bf Structural Measurements:} In this work, we focus attention on the
number of paths of distinct lengths, from $h=1$ to $3$, between each
pair of node, as well as the number of nodes reachable from each node
after $h$ steps.  This approach provides a comprehensive
characterization of the connectivity between the nodes, intrinsically
including the shortest paths.  The selection of such a comprehensive
set of measurements is justified because they reflect the existing
potential relationship between the companies.  In this work we
consider the number of paths of length $h$ between two nodes $i$ and
$j$, hence $Q_h(i,j) = Q_h(j,i)$, as the main topological superedge
measurement.

{\bf Choice of Dynamics:} The interactions between companies often
take place through pairwise communication and querying, which
typically induces chains of contact.  For instance, one company may
inquiry about a service availability to another company, which in turn
may contact another, and so on.  Naturally, such chains of contacts
avoid going through the same company more than once.  In this work we
model such contacting interactions in terms of self avoiding walks
initiating at the input set of nodes and progressing through the
network until the walk can proceed no longer or a fixed number $H$ of
steps is exceeded.  Such a dynamics properly reflects the chain of
contacts between companies as well as the avoidance of repeated
contacts.  Its main limitation is that it does not take into account
possible preferential interactions between specific companies.  Though
preferential self-avoiding walks can be implemented in principle, we
lack information about the preferential links between companies.

{\bf Dynamical Measurements:} By simulating a large number of
self-avoiding random walks initiating at the input node $i$, it is
possible to estimate the transition probabilities $P_(j,i)$,
parametrized by the path length $h$, between the input and output
($j$) nodes.  Such probabilities supply important information about
the interactions resulting from the chosen dynamics, which may or not
be related to the topological connectivity quantified by the
superedges.  Additional measurements of the outward accessibility of
the nodes complements the characterization of the dynamics of
communication between the companies with respect to the uniformity of
contacts.  More specifically, a company with high outward
accessibility $Acc\_out_{h}$ will be able to access all nodes at a
distance $h$ in the shortest period of time.  In addition, the
robustness of the communication of a company $i$ and other companies
at distance $h$ tends to increase with the accessibility of $i$.

\begin{figure}[htb]
  \vspace{0.3cm} \begin{center}
  \includegraphics[width=1\linewidth]{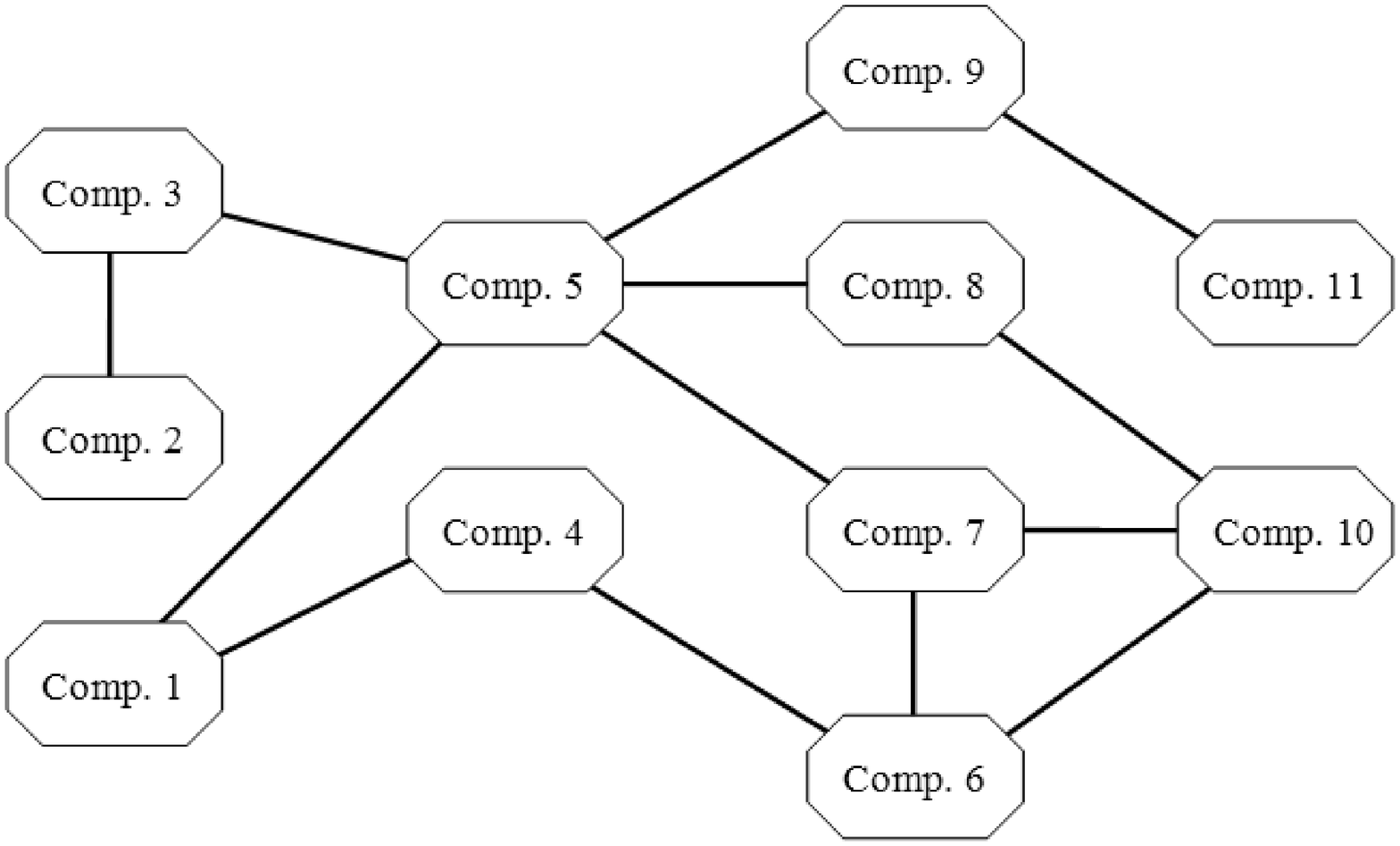} \\
  \caption{Hypothetical web of connections between tourism companies.
  }~\label{fig:exemp} \end{center}
\end{figure}

Figure~\ref{fig:exemp} presents a hypothetical relationship between 11
tourism companies, where each column corresponds to a different
\emph{region}.  The number of paths between these companies provides an
important information which is not revealed by traditional
measurements such as degrees, clustering coefficient and shortest
paths.  For instance, the fact that there are four paths (3 of length
3 and 1 of length 4) between companies 1 and 10 implies that searches
for resources only found in the latter agency are much more likely to
be well-succeeded than if the requested resource can only be found at
company 11 (only one path exists between companies 1 and 11).
Interestingly, though company 2 is linked to a single other company,
in the same region, because it has access to company 1 through the
path through companies 3 and 5, it will also have good chances of
obtaining the services from company 10.  However, a more complete
picture of the interconnectivity and probability of visits between
these companies requires the more systematic and comprehensive
characterization provided by the superedges approach.

Table~\ref{tab:ex} gives the total number of companies $T_h(i)$
reached after $h = 1, 2, 3, 4$ steps from node $i$ as well as the
inward and outward activations and accessibilities for the companies
in the hypothetical example in Figure~\ref{fig:exemp}.  We have that
the number of companies that can be reached for $h = 1, 2, 3, 4$
varies substantially for each node.  For instance node 6 reaches 3
companies at $h=2$ and 10 companies after $h=4$ steps.  However, the
number of companies that can be reached after $h$ steps does not
provide a good description of the interrelationship between the
companies because some of them may be only rarely visited.  The
situation in Figure~\ref{fig:probs}(b) illustrates this fact: though
nodes 2, 3, 4 and 5 are all reachable from node 1 after 3 steps, node
3 will be visited only sporadically ($P_3(3,1)=1/10$) compared with
the other nodes.

\begin{table*}
\caption{The number of reachable companies and inward and outward
           activations and accessibilities for $h=1, 2, 3, 4$ with
           respect to the network in Figure~\ref{fig:exemp}.}  
\vspace{0.5cm}
\begin{tabular}{||l||c|c|c|c|c|c|c|c|c|c|c||}  \hline \hline
Measurement   & 1  & 2  & 3  & 4  & 5  & 6  & 7  & 8  & 9  & 10 & 11 \\ \hline
 \hline
 $T_1$        & 2  & 1  & 2  & 2  & 5  & 3  & 3  & 2  & 2  & 3  & 1  \\ \hline
 $T_2$        & 5  & 1  & 4  & 3  & 5  & 5  & 7  & 5  & 4  & 4  & 1  \\ \hline
 $T_3$        & 5  & 4  & 4  & 6  & 5  & 5  & 7  & 7  & 4  & 7  & 4  \\ \hline
 $T_4$        & 6  & 4  & 5  & 9  & 6  & 10 & 5  & 7  & 5  & 9  & 4  \\ \hline
 \hline
 $Act\_in_1$  &0.07&0.05&0.12&0.08&0.23&0.12&0.09&0.05&0.12&0.12&0.05\\ \hline
 $Act\_in_2$  &0.08&0.02&0.05&0.05&0.32&0.12&0.13&0.08&0.05&0.08&0.02\\ \hline
 $Act\_in_3$  &0.10&0.05&0.05&0.10&0.13&0.08&0.13&0.12&0.05&0.11&0.05\\ \hline
 $Act\_in_4$  &0.08&0.05&0.03&0.10&0.12&0.14&0.06&0.06&0.03&0.13&0.05\\ \hline
 $Act\_out_1$ &1.00&1.00&1.00&1.00&1.00&1.00&1.00&1.00&1.00&1.00&1.00\\ \hline
 $Act\_out_2$ &1.00&1.00&0.50&1.00&1.00&1.00&1.00&1.00&0.50&1.00&1.00\\ \hline
 $Act\_out_3$ &1.00&1.00&0.50&1.00&1.00&1.00&1.00&1.00&0.50&1.00&1.00\\ \hline
 $Act\_out_4$ &0.75&1.00&0.38&1.00&0.60&1.00&0.83&0.75&0.38&0.96&1.00\\ \hline
 \hline
 $Acc\_in_1$  &0.13&0.12&0.14&0.13&0.20&0.15&0.14&0.12&0.14&0.15&0.12\\ \hline
 $Acc\_in_2$  &0.14&0.11&0.12&0.12&0.23&0.15&0.17&0.14&0.12&0.14&0.11\\ \hline
 $Acc\_in_3$  &0.15&0.12&0.12&0.15&0.15&0.13&0.17&0.16&0.13&0.16&0.12\\ \hline
 $Acc\_in_4$  &0.14&0.13&0.12&0.15&0.16&0.18&0.13&0.13&0.12&0.17&0.13\\ \hline
 $Acc\_out_1$ &0.20&0.10&0.20&0.20&0.50&0.30&0.30&0.20&0.20&0.30&0.10\\ \hline
 $Acc\_out_2$ &0.40&0.10&0.28&0.28&0.48&0.47&0.48&0.64&0.29&0.35&0.10\\ \hline
 $Acc\_out_3$ &0.41&0.40&0.27&0.57&0.28&0.28&0.27&0.64&0.27&0.61&0.40\\ \hline
 $Acc\_out_4$ &0.44&0.37&0.23&0.72&0.36&0.36&0.86&0.37&0.23&0.75&0.37\\ \hline
\end{tabular}~\label{tab:ex}
\end{table*}

A clearer picture of the structural and dynamical interrelationships
between each pair of companies can be provided by the superedges
approach.  From the perspective of activation, the company with the
highest inward activation (the most frequently visited) is the agency
5, with $Act\_in_2(5,\Omega) \approx 0.32$.  The least frequently
visited agencies are 3 and 9, with $Act\_in_2(3,\Omega) =
Act\_in_2(3,\Omega) = 0.03$.  Several companies are able to perform
maximum accesses (i.e. $Act\_out_h(i,j) = 1$) to other companies, but
such a capability falls steeply for companies 3 and 9, for which
$Act\_out_4(\Omega,3) = Act\_out_4(\Omega,9) = 0.38$.  Such an abrupt
decrease is a consequence of the fact that both nodes 3 and 9 have an
initial branch leading to a dead-end (edges 3 to 2 and 9 to 11).

All companies have similar inward accessibilities, except for company
5, which presents particularly high values $Acc\_in_1(i,\Omega) =
0.20$ and $Acc\_in_2(i,\Omega) = 0.23$.  The high inward activation
and accessibility of company 5 is a consequence of its `centrality'
ahd high degree in the original network.  The highest values of
outward accessibility are verified for companies 4 and 10, with
$Acc\_out_4(\Omega,4) = 0.72$ and $Acc\_out_4(\Omega,4) = 0.75$.
Therefore, these two companies are capable of effective access to a
total of 9 other companies after $h=4$ contacts.

\section{Superedges Analysis of the Elba Tourism Network}

Having described and discussed the structural and dynamical superedges
measurements, we now proceed to their application to the real-world
tourism destination of the Island of Elba.  We restrict our attention
to $h=1, 2,$ and $3$.  The isolated nodes of the original network were
not considered in the superedges investigation, leaving out 644 nodes
to be considered in our analysis.

\begin{figure*}[htb]
  \vspace{0.3cm} \begin{center}
	  \includegraphics[width=0.8\linewidth]{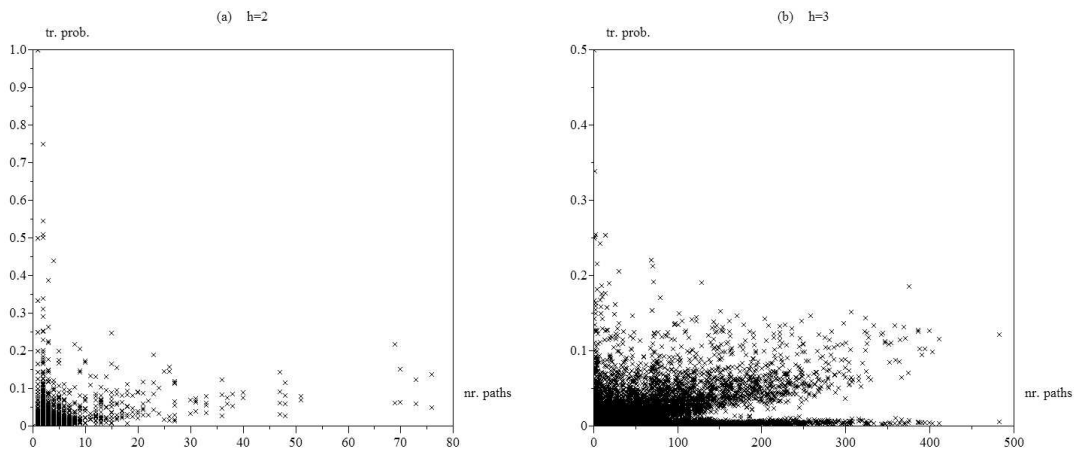} \\
  \caption{The scatterplots of the number of distinct paths and
            transition probabilities for the Elba tourism destination
            network with respect to $h=2$ (a) and $3$ (b).
  }~\label{fig:corrs} \end{center}
\end{figure*}

We start by considering possible correlations between the number of
distinct paths and the respective transition probabilities between
pairs of nodes for each $h = 2,$ and $3$ (all transitions are equal to
1 for $h=1$).  Figure~\ref{fig:corrs} shows the scatterplots obtained
for these two situations, each involving a total of 414736 edges.  The
lack of positive correlation between these pairs of
structural:dynamical measurements obtained for $h=2$
(Fig.~\ref{fig:corrs}a) makes it clear that the paths of length 2
between pairs of nodes in the Elba tourism destination network are
intensely interdependent (i.e. few triangles), implying situations
involving several paths to have small transition probabilities because
of deviations along highly interconnected paths.  A different
relationship was obtained for $h=3$ (Fig.~\ref{fig:corrs}b), which
includes two main components: a cloud of correlated points and a group
of points with high number of paths of length 3 and respective low
transition probabilities.  While the cases belonging to the former
relationship are likely to correspond to relatively independent paths
of length 3, the cases in the second group are related to intensely
interdependent paths.  Overall, the number of paths and transition
probabilities between pairs of companies varied intensely, confirming
great heterogeneity of the interconnections between the involved
companies.

\begin{figure*}[htb]
  \vspace{0.3cm} \begin{center}
	  \includegraphics[width=0.8\linewidth]{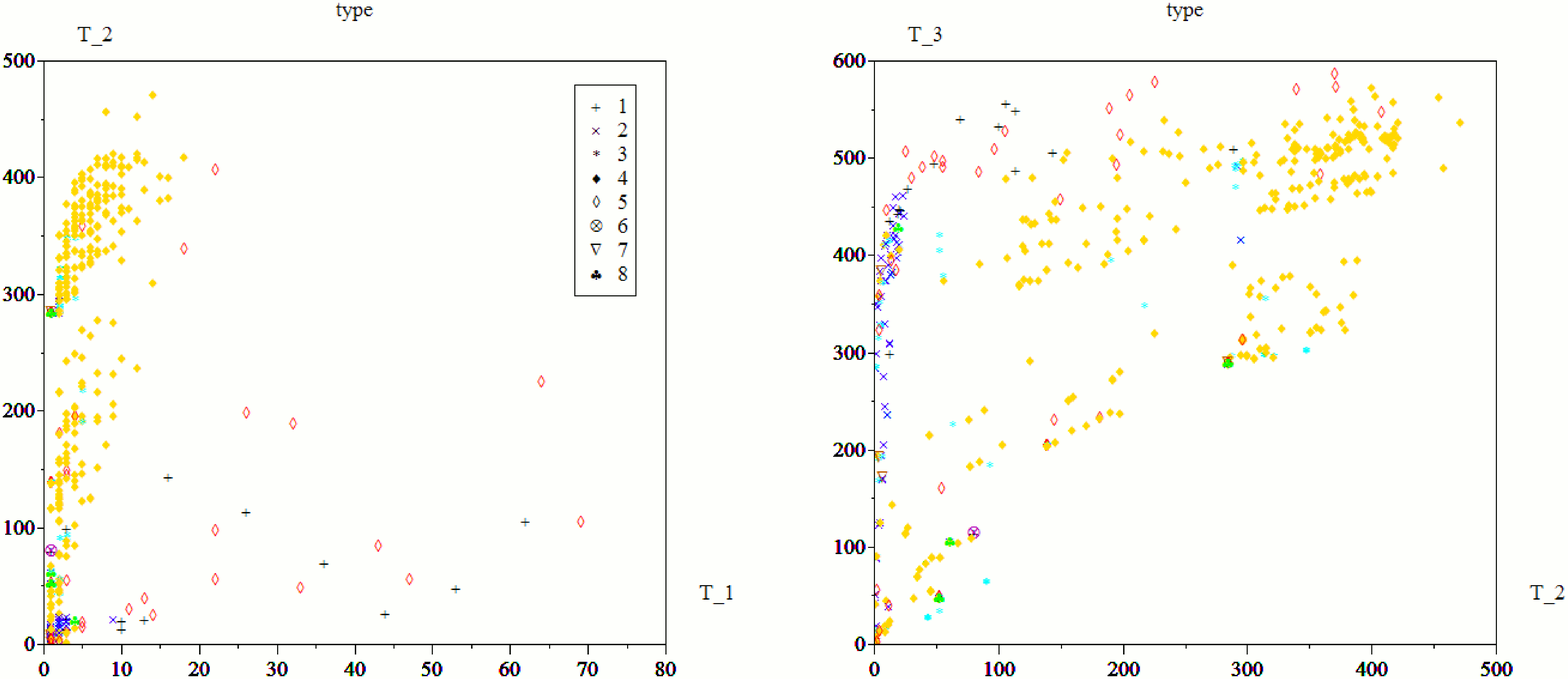} \\
	  \includegraphics[width=0.8\linewidth]{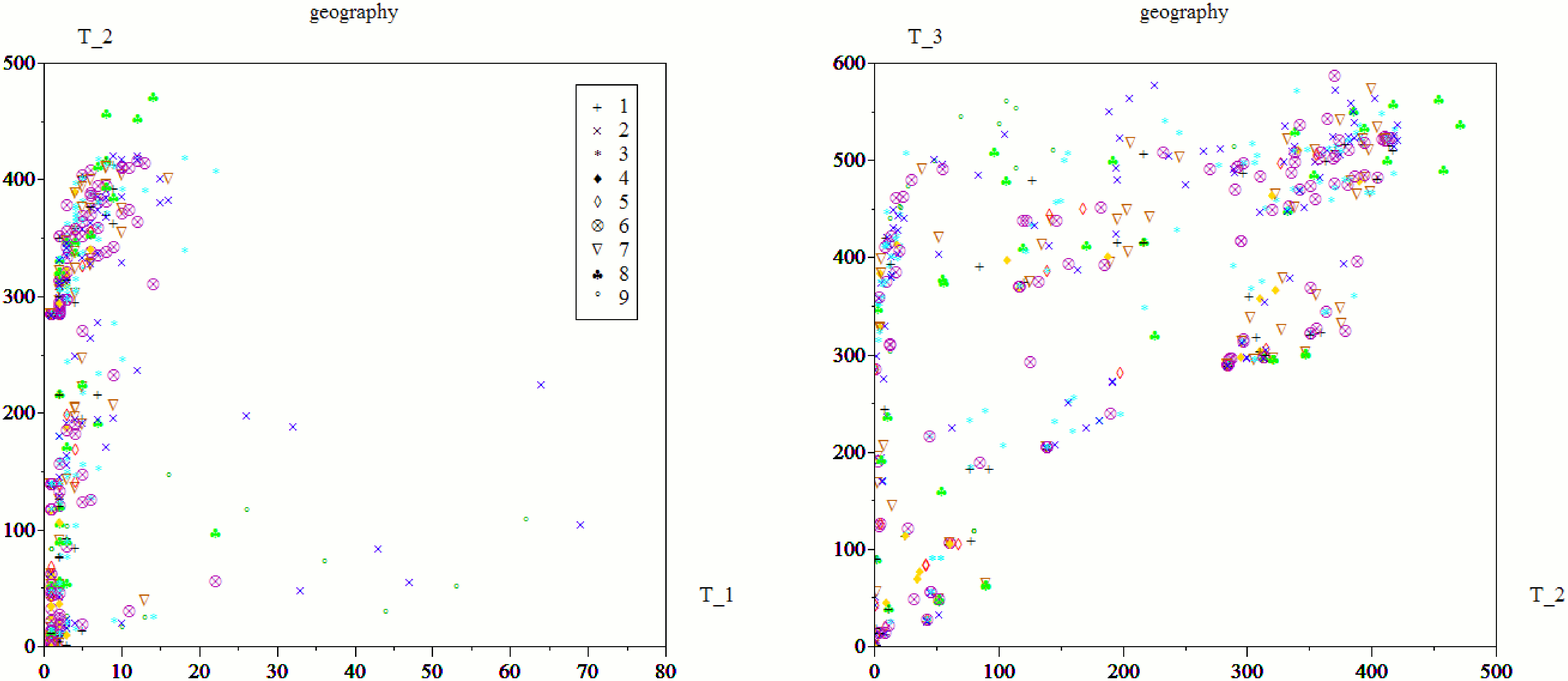} \\
	  \includegraphics[width=0.8\linewidth]{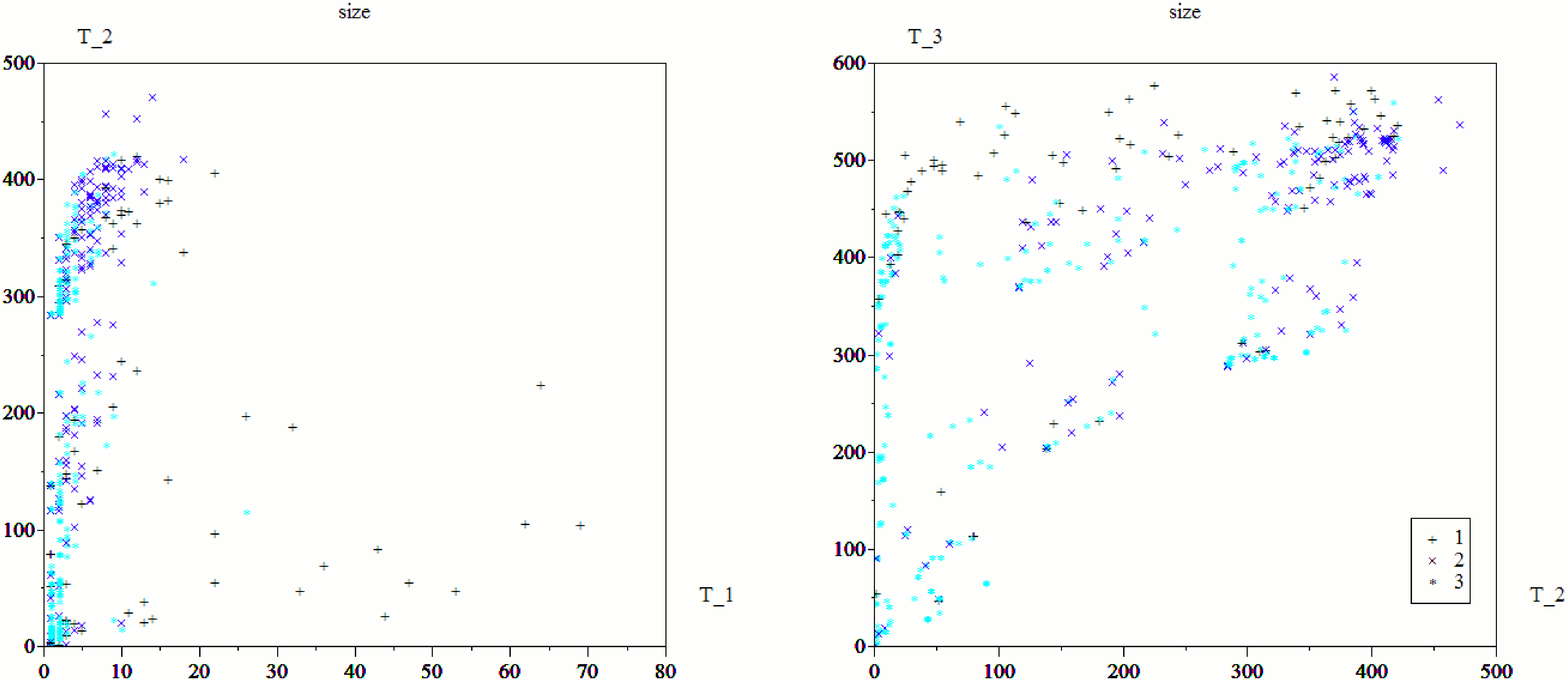} \\
  \caption{The scatterplots of the number of distinct paths for
             the three classifications of the companies, namely
             \emph{type}, \emph{geography} and \emph{size}.
  }~\label{fig:Ts} \end{center}
\end{figure*}

Figure~\ref{fig:Ts} shows the distributions of the number of reachable
companies at $h=2,$ and $3$ with respect to the type, geography and
size of the respective companies identified by the colors and symbols.
It is clear from the scatterplots identifying the types of the
companies that the number of reachable nodes at $h=2$ and $3$ depends
of the type of companies.  For instance, the companies of type 4
(hotels) tend to reach few companies for $h=1$ but can reach a large
number of companies after 2 steps.  On the other hand, companies of
types 1 and 5 (respectively associations and intermediaries) tend to
reach many companies at $h=1$.  The companies of type 1, 4 and 5 are
capable of reaching several other companies at $h=3$, while
organizations of type 2 (entertainment and cultural resources) can
reach varying numbers of companies for this same number of steps.  At
the same time, no clear trends can be inferred while considering the
geography of the networks.  The reachability is strongly dependent of
the size of the companies, with companies of size 1 (large) tending to
reach many other companies for $h=1$ and $3$.  Similar tendencies are
verified for sizes 2 (medium) and 3 (small).

\begin{figure*}[htb]
  \vspace{0.3cm} \begin{center}
	  \includegraphics[width=0.9\linewidth]{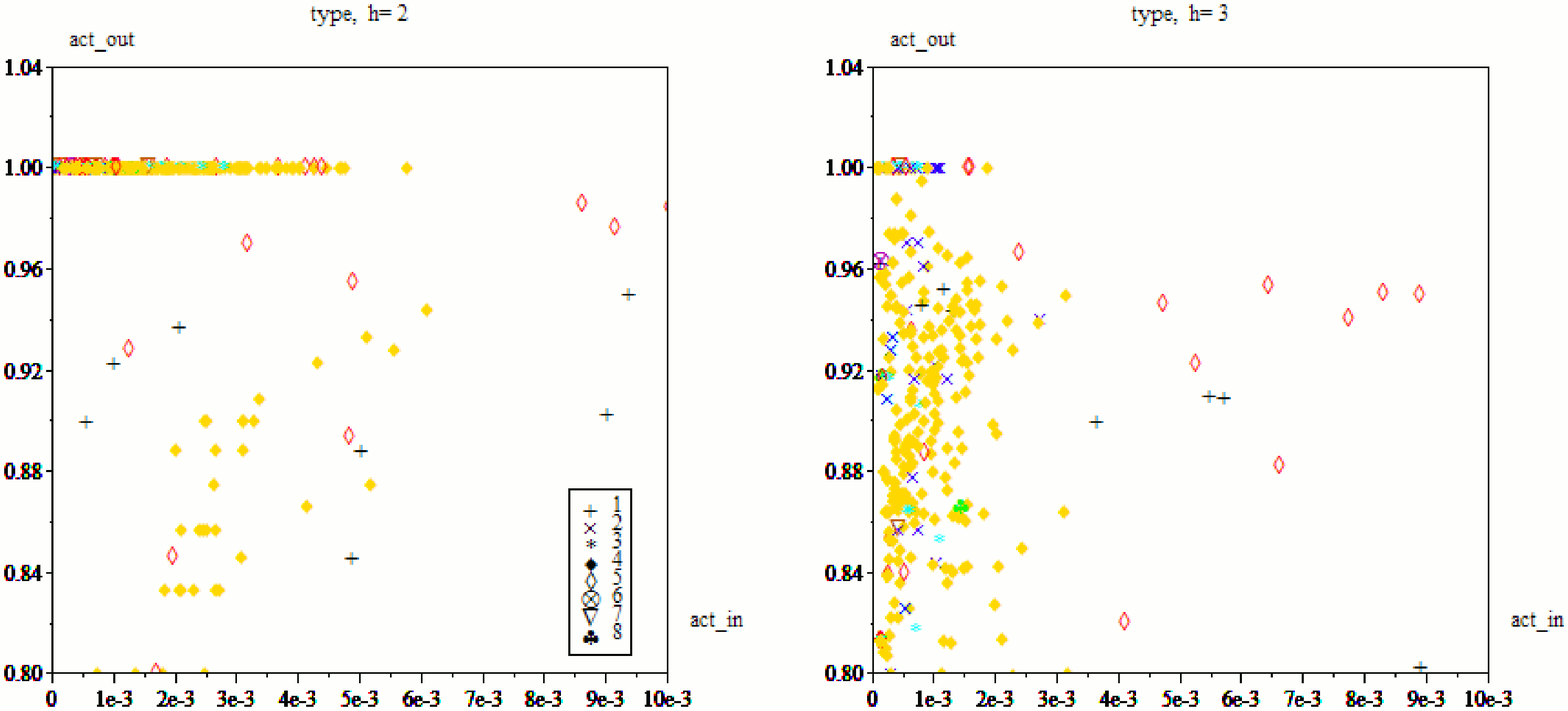} \\
	  \includegraphics[width=0.9\linewidth]{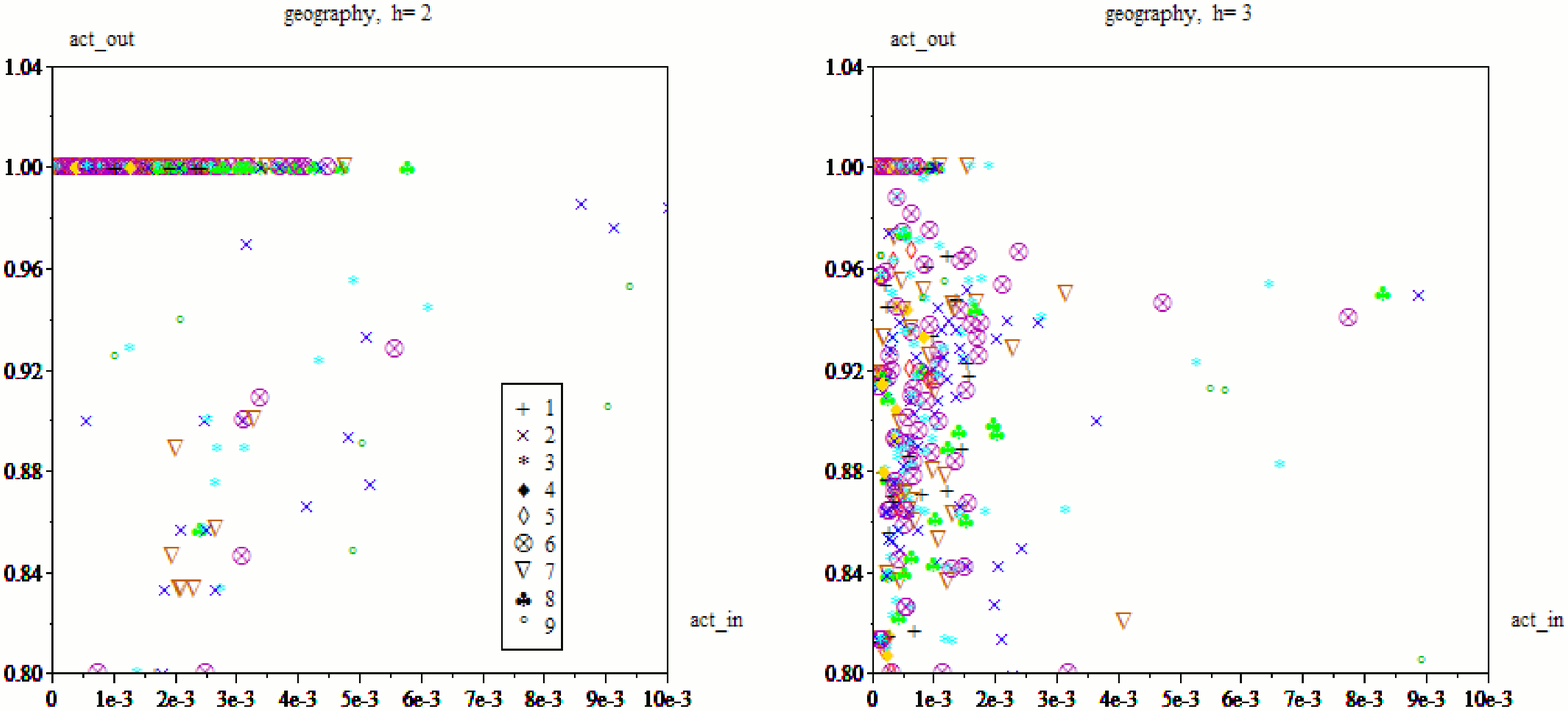} \\
	  \includegraphics[width=0.9\linewidth]{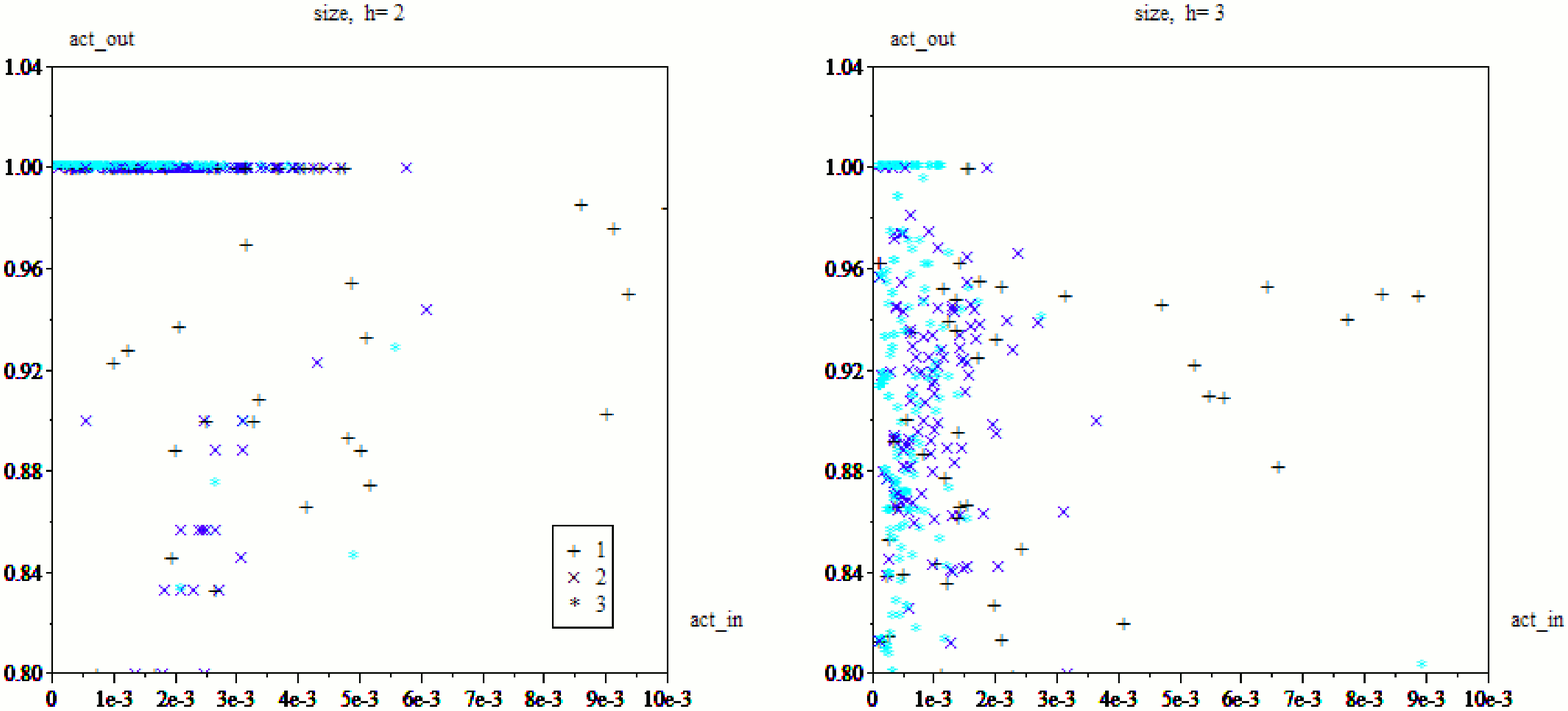} \\
  \caption{The scatterplots of the inward and outward activations for
             the three classifications of the companies, namely
             \emph{type}, \emph{geography} and \emph{size}.
  }~\label{fig:acts} \end{center}
\end{figure*}

The inward and outward activations for $h=2$ and $3$ are shown in
Figure~\ref{fig:acts}.  Again, the intrinsic activations depended
strongly on the type and size of the companies, and had little
relationship with the respective geographies.  More specifically,
companies of type 1, 4 and 5 tended to have their outward activations
reduced more intensely for $h=2$.  Most types of companies underwent
substantial reductions of outward activations for $h=3$, but companies
4 and 5 exhibited markedly distinct behaviors with respect to the
inward activation, with the former type of companies being
characterized by smaller values of inward activations.  The companies
of size 1 and 2 exhibited more intense decrease of outward activation
than companies of size 3 for $h=2$, but companies of size 1 presented
the highest inward activations for $h=3$.  This means that the latter
sizes of companies are more likely to receive queries from other
companies.

\begin{figure*}[htb]
  \vspace{0.3cm} \begin{center}
	  \includegraphics[width=1\linewidth]{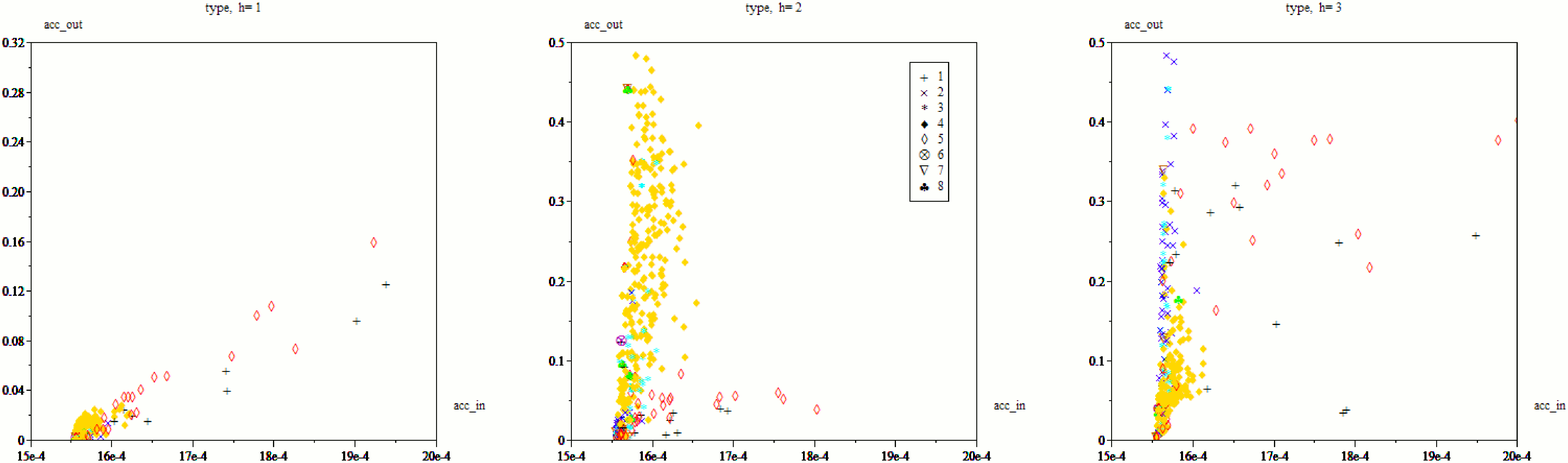} \\
	  \includegraphics[width=1\linewidth]{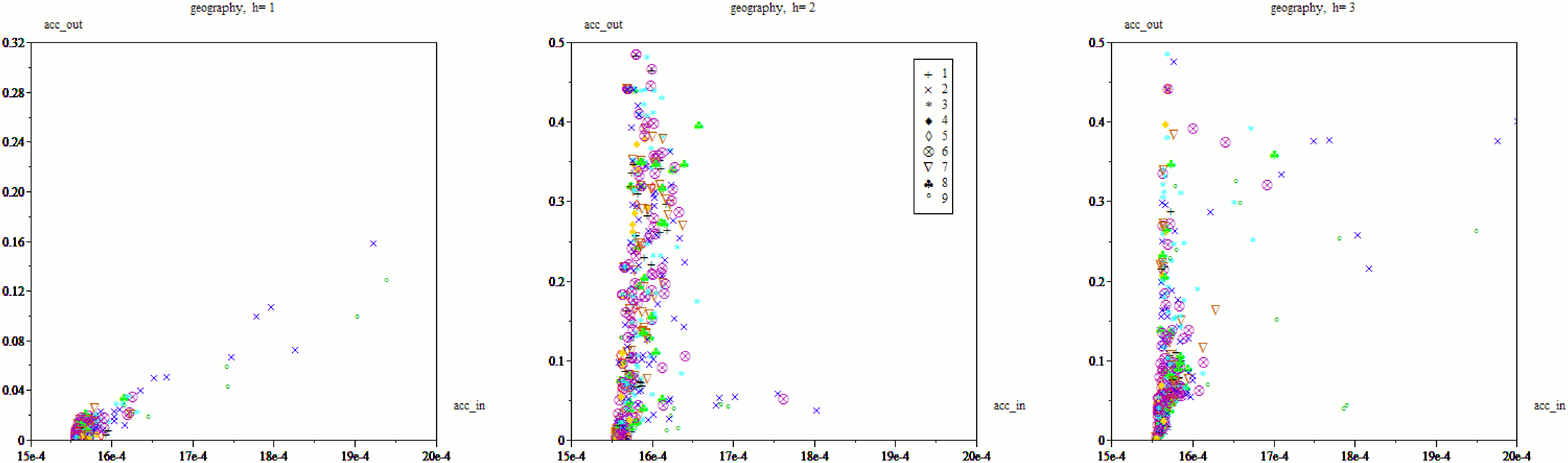} \\
	  \includegraphics[width=1\linewidth]{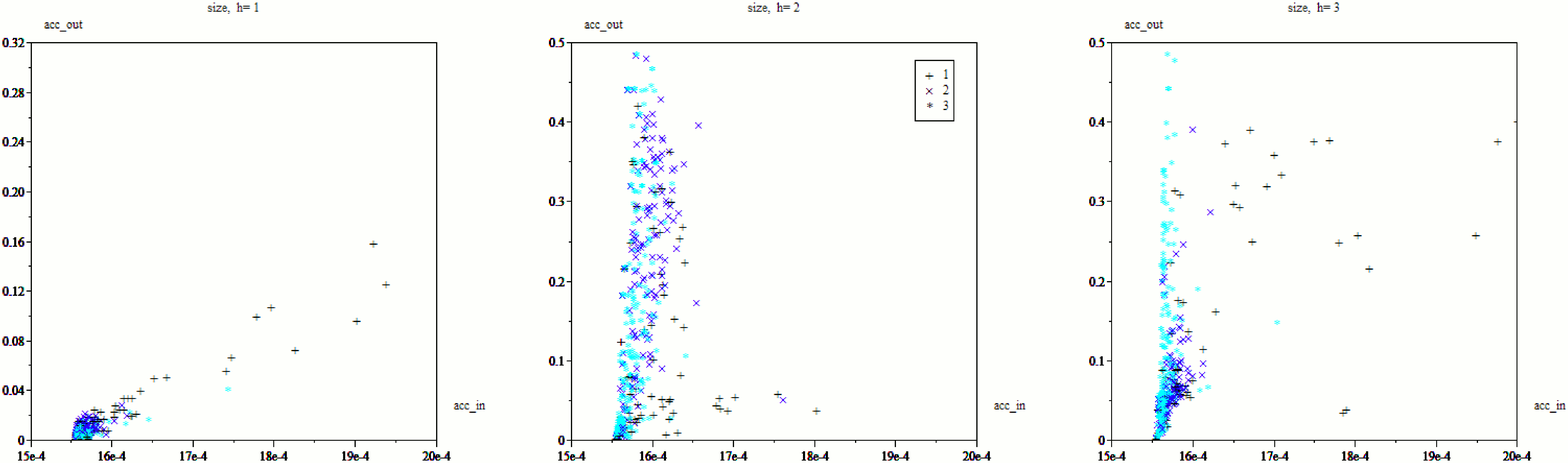} \\
  \caption{The scatterplots of the inward and outward accessibilities for
             the three classifications of the companies, namely
             \emph{type}, \emph{geography} and \emph{size}.
  }~\label{fig:accs} \end{center}
\end{figure*}

Figure~\ref{fig:accs} depicts the inward and outward accessibilities
for each of the three classification schemes of companies, namely by
type, geography and size, with respect to $h=1, 2,$ and $3$.  No clear
tendencies are observed regarding the geographic types, but the inward
and outward accessibilities are strongly related to the type and size
of the involved companies.  More specifically, companies of type 2, 3,
4 and 5 presented rather distinct accessibilities for all values of
$h$, tending to be characterized by higher inward accessibility.  This
means that this type of companies is reached with higher efficiency by
several other companies, i.e. each of these companies will be accessed
more quickly by queries originating at the respective reachable
companies.  As observed for the number of reachable companies and
inward/outward activations, the companies of size 1 tend to have a
markedly distinct behavior also with respect to their inward/outward
accessibilities.  More specifically, this type of companies tend to be
visited in a brief period of time by queries originating at the
reachable companies.

All in all, in addition to the verified specific trends, the above
results make it clear that the structural and dynamical properties of
the considered companies depend strongly on their type and size, being
much less affected by their respective geographical position.

By giving a 'physical' dynamical interpretation to these results, it
is possible to summarize the outcomes as follows. The geographical
subdivision does not seem to have any sort of influence on the
reachability of Elban tourism companies. This is in agreement with the
observed very low tendency to form 'geographic' communities coming
from the static modularity analysis.  Some companies or organizations
(mainly associations and intermediaries) seem to be very active in
reaching or being reached by other companies. The size of the company
or organization matters, as larger entities tend to be more active than
the smaller ones.  This picture is in full agreement of what is known
of the behaviors of tourism operators in the destination (see for
example~\cite{Tallinucci:2006} or~\cite{Pechlaner:2003}).  However,
the identification of the specific trends exhibited by each type of
company only became clearer through the superedges integration of the
structural and dynamical features.

\section{Concluding Remarks}

Tourism destination networks are amongst the most complex real-world
systems, involving intrincated structure and non-linear dynamics.
Because of the economical importance of such systems, it becomes
increasingly important to devise effective means for describing,
characterizing and modeling these systems so that their structure and
dynamics are better understood, allowing predictions, identification
of possible improvements, and simulations aimed at evaluating the
effects of varying scenarios and conditions.  At the same time,
complex networks research is now a mature area catering for all such
requirements, from the characterization of the structure of the
interaction between tourism companies to the relationship with the
respective dynamics of interactions and information exchange.  The
current article has brought these two important areas together with
respect to the comprehensive analysis of the Elba tourism destination
network, a real-world structure which has been recently obtained
through systematic and careful investigation including field data
collection.  The main contributions of our work are listed and
discussed as follows:

\emph{Complete Real-World Example of the Superedges Approach:} This
work represents the very first comprehensive application of the
superedges approach to a real-world network.  As such, special care
has been invested in order to making the respective concepts and
interpretations clear from the context of the specific application,
namely tourism destination structures.  In particular, each of the
choices which have to be made regarding the superedges methodology ---
including input/output, structural measurements, dynamics and
dynamical measurements --- have been motivated and justified with
respect to the tourism application.  The methodology was first
illustrated with respect to a hypothetical simple network of
companies, and then applied systematically to the real-world Elba
tourism destination structure.  As such, this first complete example
can be used as a application reference guide for researchers intending
to apply the superedges approach to other specific problems.

\emph{Comprehensive Structural Characterization:} This work has
presented one of the very first comprehensive analysis of the
topological characteristics of a tourism destination network,
including traditional measurements, modularity as well as the number
of distinct paths between pairs of nodes. A comparison, and the
substantial similarity, of the outcomes with previous knowledge of the
relationships among the tourism companies located at Elba
stakeholders~\cite{Pechlaner:2003, Tallinucci:2006} substantiates the
effectiveness of our approach and results.

\emph{Comprehensive Dynamical Characterization:} In addition to
characterizing several structural properties of the tourism
destination network, we performed a systematic investigation of a
possible model of the dynamics of interactions between the involved
companies, which was done by considering self-avoiding random walks.
The effects of such a dynamics over the interaction between the
companies has been effectively expressed in terms of the inward and
outward activations and accessibilities.

\emph{Practical Implications:}  As for the static structural 
characterization, the results of the analysis conducted by using the
dynamic superedges approach finds a justification and a verification
in the established (qualitative) knowledge of the destination
examined, its stakeholders and their behavior.  This further confirms
and reinforces the validity of the currently adopted
approach. Although intriguing per se, the outcomes may have, at a
destination management level even a bigger importance.  A quantitative
investigation method, with a strong theoretical basis, is able to
provide descriptions and indications which, traditionally would have
required long, and sometimes disputable, qualitative studies.  The
reliability of the conclusions one may find in combining the two
approaches (qualitative and quantitative) can stand any
comparison. The combination can be used in several ways, but
principally in confirming or correcting a previous knowledge or the
one coming form more traditional social studies.  In our case, for
example, we have seen it would be possible to improve the overall
collaborative environment by considering our destination as a single
'geographical entity', disregarding any pre-set administrative
division that, on the contrary, is the one normally used as a basis
for the design of planning and policy actions.  We have also
identified, in an undisputable way, the stakeholders of this
destination which would require the highest effort. It is possible, in
other words, to reliably assign priorities to plans and actions and to
distribute more productively resources (typically scarce) for their
implementation.  As many scholars in the field of tourism know, this
is a crucial issue for an efficient and effective management of a
destination and for favoring its socio-economic
growth~\cite{Ritchie:2003, Vanhove:2005}.

The future works implied by our currently described research include
but are not limited to the following possibilities:

\emph{Analysis of Other Destination Networks:} It would be particularly
interesting to apply the reported methodology to other destination
networks, in order to allow comparisons between the respective
structures and dynamics.  Among the several related possibilities, it
would be interesting to compare the Elba network with networks
obtained for tourism regions in other continents, such as America and
Asia, in order to search for similar and distinct properties.  The
reported methodology can also be applied as a means to obtain a
comparative analysis between tourism destinations in the first and
third world.

\emph{Simulations:} Given a network such as that analyzed in this work
and its respective comprehensive characterization, it would be
interesting to perform simulations involving the addition or removal
of connections, in order to investigate effects of communication
failures as well as the creation of new partnerships.  

\emph{Inference of Growth Models:} The comprehensive characterization
of the Elba tourism destination network in terms of structural and
dynamical features reported in this article has paved the way to
attempts to obtain growth models capable of reproducing the observed
structure, possibly under the effect of the respective dynamics.  For
instance, after starting with a small random structure, new
connections could be established while taking into account regional
proximity, possibly involving the path-regular knitted
network~\cite{Costa_comp:2007} to interconnect the companies according
to their positions.  Such growth models would allow a yet more
complete understanding of tourism destination systems.

\begin{acknowledgments}
Luciano da F. Costa thanks CNPq (308231/03-1) and FAPESP (05/00587-5)
for sponsorship.
\end{acknowledgments}

\bibliography{elba2}

\begin{thebibliography}{33}
\expandafter\ifx\csname natexlab\endcsname\relax\def\natexlab#1{#1}\fi
\expandafter\ifx\csname bibnamefont\endcsname\relax
  \def\bibnamefont#1{#1}\fi
\expandafter\ifx\csname bibfnamefont\endcsname\relax
  \def\bibfnamefont#1{#1}\fi
\expandafter\ifx\csname citenamefont\endcsname\relax
  \def\citenamefont#1{#1}\fi
\expandafter\ifx\csname url\endcsname\relax
  \def\url#1{\texttt{#1}}\fi
\expandafter\ifx\csname urlprefix\endcsname\relax\def\urlprefix{URL }\fi
\providecommand{\bibinfo}[2]{#2}
\providecommand{\eprint}[2][]{\url{#2}}

\bibitem[{\citenamefont{Vanhove}(2005)}]{Vanhove:2005}
\bibinfo{author}{\bibfnamefont{N.}~\bibnamefont{Vanhove}},
  \emph{\bibinfo{title}{Economics of Tourism Destination}}
  (\bibinfo{publisher}{Elsevier Butterworth-Heinemann}, \bibinfo{year}{2005}).

\bibitem[{\citenamefont{Jafari}(2000)}]{Jafari:2000}
\bibinfo{author}{\bibfnamefont{J.}~\bibnamefont{Jafari}},
  \emph{\bibinfo{title}{Encyclopaedia of Tourism}}
  (\bibinfo{publisher}{Routledge}, \bibinfo{year}{2000}).

\bibitem[{\citenamefont{Baggio}(2008)}]{Baggio:2008}
\bibinfo{author}{\bibfnamefont{R.}~\bibnamefont{Baggio}}
  (\bibinfo{year}{2008}), \bibinfo{note}{arXiv:physics/0701063}.

\bibitem[{\citenamefont{Faulkner and Russell}(1997)}]{Faulkner:1997}
\bibinfo{author}{\bibfnamefont{B.}~\bibnamefont{Faulkner}} \bibnamefont{and}
  \bibinfo{author}{\bibfnamefont{R.}~\bibnamefont{Russell}},
  \bibinfo{journal}{Pacif. Tour. Rev.} \textbf{\bibinfo{volume}{1}},
  \bibinfo{pages}{93} (\bibinfo{year}{1997}).

\bibitem[{\citenamefont{Lazzeretti and Petrillo}(2003)}]{Lazzeretti:2006}
\bibinfo{author}{\bibfnamefont{L.}~\bibnamefont{Lazzeretti}} \bibnamefont{and}
  \bibinfo{author}{\bibfnamefont{C.~S.} \bibnamefont{Petrillo}},
  \emph{\bibinfo{title}{Tourism Local Systems and Networks}}
  (\bibinfo{publisher}{Elsevier}, \bibinfo{year}{2003}).

\bibitem[{\citenamefont{Morrison et~al.}(2004)\citenamefont{Morrison, Lynch,
  and Johns}}]{Morrison:2004}
\bibinfo{author}{\bibfnamefont{A.}~\bibnamefont{Morrison}},
  \bibinfo{author}{\bibfnamefont{P.}~\bibnamefont{Lynch}}, \bibnamefont{and}
  \bibinfo{author}{\bibfnamefont{N.}~\bibnamefont{Johns}},
  \bibinfo{journal}{Intl. J. Contemp. Hosp. Manag.}
  \textbf{\bibinfo{volume}{16}}, \bibinfo{pages}{197} (\bibinfo{year}{2004}).

\bibitem[{\citenamefont{Pavlovich}(2003)}]{Pavlovich:2003}
\bibinfo{author}{\bibfnamefont{K.}~\bibnamefont{Pavlovich}},
  \bibinfo{journal}{Tour. Manag.} \textbf{\bibinfo{volume}{24}},
  \bibinfo{pages}{203} (\bibinfo{year}{2003}).

\bibitem[{\citenamefont{Scott et~al.}(2008{\natexlab{a}})\citenamefont{Scott,
  Cooper, and Baggio}}]{Scott:2008}
\bibinfo{author}{\bibfnamefont{N.}~\bibnamefont{Scott}},
  \bibinfo{author}{\bibfnamefont{C.}~\bibnamefont{Cooper}}, \bibnamefont{and}
  \bibinfo{author}{\bibfnamefont{R.}~\bibnamefont{Baggio}},
  \bibinfo{journal}{Anns. Tour. Res.} \textbf{\bibinfo{volume}{35}},
  \bibinfo{pages}{169} (\bibinfo{year}{2008}{\natexlab{a}}).

\bibitem[{\citenamefont{Scott et~al.}(2008{\natexlab{b}})\citenamefont{Scott,
  Baggio, and Cooper}}]{Scott_b:2008}
\bibinfo{author}{\bibfnamefont{N.}~\bibnamefont{Scott}},
  \bibinfo{author}{\bibfnamefont{R.}~\bibnamefont{Baggio}}, \bibnamefont{and}
  \bibinfo{author}{\bibfnamefont{C.}~\bibnamefont{Cooper}},
  \emph{\bibinfo{title}{Network Analysis and Tourism: From Theory to Practice}}
  (\bibinfo{publisher}{Channel View}, \bibinfo{year}{2008}{\natexlab{b}}).

\bibitem[{\citenamefont{Tallinucci and Testa}(2006)}]{Tallinucci:2006}
\bibinfo{author}{\bibfnamefont{V.}~\bibnamefont{Tallinucci}} \bibnamefont{and}
  \bibinfo{author}{\bibfnamefont{M.}~\bibnamefont{Testa}},
  \emph{\bibinfo{title}{Marketing per le isole}} (\bibinfo{publisher}{Franco
  Angeli}, \bibinfo{address}{Milano}, \bibinfo{year}{2006}).

\bibitem[{\citenamefont{Bramwell and Lane}(2000)}]{Bramwell:2000}
\bibinfo{author}{\bibfnamefont{B.}~\bibnamefont{Bramwell}} \bibnamefont{and}
  \bibinfo{author}{\bibfnamefont{B.}~\bibnamefont{Lane}},
  \emph{\bibinfo{title}{Tourism Collaboration and Partnerships: Politics
  Practice and Sustainability}} (\bibinfo{publisher}{Channel View},
  \bibinfo{year}{2000}).

\bibitem[{\citenamefont{Baggio}(2007)}]{Baggio:2007}
\bibinfo{author}{\bibfnamefont{R.}~\bibnamefont{Baggio}},
  \bibinfo{journal}{Phys. A} \textbf{\bibinfo{volume}{379}},
  \bibinfo{pages}{727} (\bibinfo{year}{2007}).

\bibitem[{\citenamefont{Cooper}(2006)}]{Cooper:2006}
\bibinfo{author}{\bibfnamefont{C.}~\bibnamefont{Cooper}},
  \bibinfo{journal}{Ann. Tour. Res.} \textbf{\bibinfo{volume}{33}},
  \bibinfo{pages}{47} (\bibinfo{year}{2006}).

\bibitem[{\citenamefont{Scott et~al.}(2007)\citenamefont{Scott, Cooper, and
  Baggio}}]{Scott:2007}
\bibinfo{author}{\bibfnamefont{N.}~\bibnamefont{Scott}},
  \bibinfo{author}{\bibfnamefont{C.}~\bibnamefont{Cooper}}, \bibnamefont{and}
  \bibinfo{author}{\bibfnamefont{R.}~\bibnamefont{Baggio}}, in
  \emph{\bibinfo{booktitle}{Advances in Tourism Marketing Conference}}
  (\bibinfo{address}{Valencia, Spain}, \bibinfo{year}{2007}).

\bibitem[{\citenamefont{Flory}(1941)}]{Flory}
\bibinfo{author}{\bibfnamefont{P.~J.} \bibnamefont{Flory}},
  \bibinfo{journal}{Journal of the American Chemical Society}
  \textbf{\bibinfo{volume}{63}}, \bibinfo{pages}{3083} (\bibinfo{year}{1941}).

\bibitem[{\citenamefont{Erd\H{o}s and R\'enyi}(1959)}]{Erdos_Reny:1959}
\bibinfo{author}{\bibfnamefont{P.}~\bibnamefont{Erd\H{o}s}} \bibnamefont{and}
  \bibinfo{author}{\bibfnamefont{A.}~\bibnamefont{R\'enyi}},
  \bibinfo{journal}{Publicationes Mathematicae (Debrecen)}
  \textbf{\bibinfo{volume}{6}}, \bibinfo{pages}{290} (\bibinfo{year}{1959}).

\bibitem[{\citenamefont{Albert and Barab\'asi}(2002)}]{Albert_Barab:2002}
\bibinfo{author}{\bibfnamefont{R.}~\bibnamefont{Albert}} \bibnamefont{and}
  \bibinfo{author}{\bibfnamefont{A.~L.} \bibnamefont{Barab\'asi}},
  \bibinfo{journal}{Rev. Mod. Phys.} \textbf{\bibinfo{volume}{74}},
  \bibinfo{pages}{47} (\bibinfo{year}{2002}).

\bibitem[{\citenamefont{Newman}(2003)}]{Newman:2003}
\bibinfo{author}{\bibfnamefont{M.~E.~J.} \bibnamefont{Newman}},
  \bibinfo{journal}{SIAM Rev.} \textbf{\bibinfo{volume}{45}},
  \bibinfo{pages}{167} (\bibinfo{year}{2003}).

\bibitem[{\citenamefont{da~F.~Costa
  et~al.}(2007{\natexlab{a}})\citenamefont{da~F.~Costa, Rodrigues, Travieso,
  and Boas}}]{Costa_surv:2007}
\bibinfo{author}{\bibfnamefont{L.}~\bibnamefont{da~F.~Costa}},
  \bibinfo{author}{\bibfnamefont{F.~A.} \bibnamefont{Rodrigues}},
  \bibinfo{author}{\bibfnamefont{G.}~\bibnamefont{Travieso}}, \bibnamefont{and}
  \bibinfo{author}{\bibfnamefont{P.~R.~V.} \bibnamefont{Boas}},
  \bibinfo{journal}{Advs. in Phys.} \textbf{\bibinfo{volume}{56}},
  \bibinfo{pages}{167} (\bibinfo{year}{2007}{\natexlab{a}}).

\bibitem[{\citenamefont{Dorogovtsev and Mendes}(2002)}]{Dorogov_Mendes:2002}
\bibinfo{author}{\bibfnamefont{S.~N.} \bibnamefont{Dorogovtsev}}
  \bibnamefont{and} \bibinfo{author}{\bibfnamefont{J.~F.~F.}
  \bibnamefont{Mendes}}, \bibinfo{journal}{Advs. in Phys.}
  \textbf{\bibinfo{volume}{51}}, \bibinfo{pages}{1079} (\bibinfo{year}{2002}).

\bibitem[{\citenamefont{Boccaletti et~al.}(2006)\citenamefont{Boccaletti,
  Latora, Moreno, Chavez, and Hwang}}]{Boccaletti:2006}
\bibinfo{author}{\bibfnamefont{S.}~\bibnamefont{Boccaletti}},
  \bibinfo{author}{\bibfnamefont{V.}~\bibnamefont{Latora}},
  \bibinfo{author}{\bibfnamefont{Y.}~\bibnamefont{Moreno}},
  \bibinfo{author}{\bibfnamefont{M.}~\bibnamefont{Chavez}}, \bibnamefont{and}
  \bibinfo{author}{\bibfnamefont{D.}~\bibnamefont{Hwang}},
  \bibinfo{journal}{Phys. Rep.} \textbf{\bibinfo{volume}{424}},
  \bibinfo{pages}{175} (\bibinfo{year}{2006}).

\bibitem[{\citenamefont{da~F.~Costa
  et~al.}(2007{\natexlab{b}})\citenamefont{da~F.~Costa, {Oliveira Jr},
  Travieso, Rodrigues, Boas, Antiqueira, Viana, and {da
  Rocha}}}]{Costa_appl:2008}
\bibinfo{author}{\bibfnamefont{L.}~\bibnamefont{da~F.~Costa}},
  \bibinfo{author}{\bibfnamefont{O.~N.} \bibnamefont{{Oliveira Jr}}},
  \bibinfo{author}{\bibfnamefont{G.}~\bibnamefont{Travieso}},
  \bibinfo{author}{\bibfnamefont{F.~A.} \bibnamefont{Rodrigues}},
  \bibinfo{author}{\bibfnamefont{P.~R.~V.} \bibnamefont{Boas}},
  \bibinfo{author}{\bibfnamefont{L.}~\bibnamefont{Antiqueira}},
  \bibinfo{author}{\bibfnamefont{M.~P.} \bibnamefont{Viana}}, \bibnamefont{and}
  \bibinfo{author}{\bibfnamefont{L.~E.~C.} \bibnamefont{{da Rocha}}}
  (\bibinfo{year}{2007}{\natexlab{b}}), \bibinfo{note}{arXiv:0711.3199}.

\bibitem[{\citenamefont{Jeong et~al.}(2001)\citenamefont{Jeong, Mason,
  Barab\'asi, and Oltvai}}]{Jeong:2001}
\bibinfo{author}{\bibfnamefont{H.}~\bibnamefont{Jeong}},
  \bibinfo{author}{\bibfnamefont{S.~P.} \bibnamefont{Mason}},
  \bibinfo{author}{\bibfnamefont{A.~L.} \bibnamefont{Barab\'asi}},
  \bibnamefont{and} \bibinfo{author}{\bibfnamefont{Z.~N.}
  \bibnamefont{Oltvai}}, \bibinfo{journal}{Nature}
  \textbf{\bibinfo{volume}{411}}, \bibinfo{pages}{41} (\bibinfo{year}{2001}).

\bibitem[{\citenamefont{Lehmann et~al.}(2003)\citenamefont{Lehmann, Lautrup,
  and Jackson}}]{Lehmann:2003}
\bibinfo{author}{\bibfnamefont{S.}~\bibnamefont{Lehmann}},
  \bibinfo{author}{\bibfnamefont{B.}~\bibnamefont{Lautrup}}, \bibnamefont{and}
  \bibinfo{author}{\bibfnamefont{A.~D.} \bibnamefont{Jackson}},
  \bibinfo{journal}{Phys. Rev. E} \textbf{\bibinfo{volume}{68}},
  \bibinfo{pages}{026113} (\bibinfo{year}{2003}).

\bibitem[{\citenamefont{da~F.~Costa}(2008)}]{Costa_superedges:2008}
\bibinfo{author}{\bibfnamefont{L.}~\bibnamefont{da~F.~Costa}}
  (\bibinfo{year}{2008}), \bibinfo{note}{arXiv:0801.4068}.

\bibitem[{\citenamefont{Pechlaner et~al.}(2003)\citenamefont{Pechlaner,
  Tallinucci, Abfalter, and Rienzner}}]{Pechlaner:2003}
\bibinfo{author}{\bibfnamefont{H.}~\bibnamefont{Pechlaner}},
  \bibinfo{author}{\bibfnamefont{V.}~\bibnamefont{Tallinucci}},
  \bibinfo{author}{\bibfnamefont{D.}~\bibnamefont{Abfalter}}, \bibnamefont{and}
  \bibinfo{author}{\bibfnamefont{H.}~\bibnamefont{Rienzner}}, in
  \emph{\bibinfo{booktitle}{Information and Communication Technologies in
  Tourism}}, edited by \bibinfo{editor}{\bibfnamefont{A.~J.}
  \bibnamefont{Frew}}, \bibinfo{editor}{\bibfnamefont{M.}~\bibnamefont{Hitz}},
  \bibnamefont{and} \bibinfo{editor}{\bibfnamefont{P.}~\bibnamefont{O'Connor}}
  (\bibinfo{publisher}{Springer}, \bibinfo{address}{Wien},
  \bibinfo{year}{2003}), pp. \bibinfo{pages}{105--114}.

\bibitem[{\citenamefont{Olsen}(2004)}]{Olsen:2004}
\bibinfo{author}{\bibfnamefont{W.}~\bibnamefont{Olsen}}, in
  \emph{\bibinfo{booktitle}{Developments in Sociology: An Annual Review}},
  edited by \bibinfo{editor}{\bibfnamefont{M.}~\bibnamefont{Holborn}}
  (\bibinfo{publisher}{Causeway Press}, \bibinfo{address}{Ormskirk, UK},
  \bibinfo{year}{2004}).

\bibitem[{\citenamefont{Clauset et~al.}(2007)\citenamefont{Clauset, Shalizi,
  and Newman}}]{Clauset:2007}
\bibinfo{author}{\bibfnamefont{A.}~\bibnamefont{Clauset}},
  \bibinfo{author}{\bibfnamefont{C.~R.} \bibnamefont{Shalizi}},
  \bibnamefont{and} \bibinfo{author}{\bibfnamefont{M.~E.~J.}
  \bibnamefont{Newman}} (\bibinfo{year}{2007}),
  \bibinfo{note}{arXiv:physics/0706.1062}.

\bibitem[{\citenamefont{Newman and Girvan}(2004)}]{Newman_Girvan:2004}
\bibinfo{author}{\bibfnamefont{M.~E.~J.} \bibnamefont{Newman}}
  \bibnamefont{and} \bibinfo{author}{\bibfnamefont{M.}~\bibnamefont{Girvan}},
  \bibinfo{journal}{Phys. Rev. E} \textbf{\bibinfo{volume}{69}},
  \bibinfo{pages}{26113} (\bibinfo{year}{2004}).

\bibitem[{\citenamefont{Clauset et~al.}(2004)\citenamefont{Clauset, Newman, and
  Moore}}]{Clauset:2004}
\bibinfo{author}{\bibfnamefont{A.}~\bibnamefont{Clauset}},
  \bibinfo{author}{\bibfnamefont{M.~E.~J.} \bibnamefont{Newman}},
  \bibnamefont{and} \bibinfo{author}{\bibfnamefont{C.}~\bibnamefont{Moore}},
  \bibinfo{journal}{Phys. Rev. E} \textbf{\bibinfo{volume}{70}},
  \bibinfo{pages}{066111} (\bibinfo{year}{2004}).

\bibitem[{\citenamefont{Fortunato and Barth\'elemy}(2007)}]{Fortunato:2007}
\bibinfo{author}{\bibfnamefont{S.}~\bibnamefont{Fortunato}} \bibnamefont{and}
  \bibinfo{author}{\bibfnamefont{M.}~\bibnamefont{Barth\'elemy}},
  \bibinfo{journal}{Proc. Natl. Acad. Sci. USA} \textbf{\bibinfo{volume}{104}},
  \bibinfo{pages}{36–41} (\bibinfo{year}{2007}).

\bibitem[{\citenamefont{Ritchie and Crouch}(2003)}]{Ritchie:2003}
\bibinfo{author}{\bibfnamefont{J.~R.~B.} \bibnamefont{Ritchie}}
  \bibnamefont{and} \bibinfo{author}{\bibfnamefont{G.~I.}
  \bibnamefont{Crouch}}, \emph{\bibinfo{title}{The Competitive Destination: A
  Sustainable Tourism Perspective}} (\bibinfo{publisher}{CABI Publishing},
  \bibinfo{year}{2003}).

\bibitem[{\citenamefont{da~F.~Costa}(2007)}]{Costa_comp:2007}
\bibinfo{author}{\bibfnamefont{L.}~\bibnamefont{da~F.~Costa}}
  (\bibinfo{year}{2007}), \bibinfo{note}{arXiv:0711.2736}.

\end{thebibliography}
\end{document}